\documentclass{ieeeaccess}
\usepackage[pdftex]{graphicx} 
\usepackage[T1]{fontenc}
\usepackage{cite}

\usepackage{amsmath,amssymb,amsfonts}
\usepackage{algpseudocode, algorithm}
\usepackage{braket}

\usepackage[hyphens]{url}

\usepackage[font=footnotesize]{subcaption}

\usepackage{soul}
\usepackage{hyperref}
\hypersetup{colorlinks=false}

\def\BibTeX{{\rm B\kern-.05em{\sc i\kern-.025em b}\kern-.08em
    T\kern-.1667em\lower.7ex\hbox{E}\kern-.125emX}}
\begin{document}
\history{}
\doi{10.1109/tqe.2025.3588783}

\title{Fidelity-Aware Multipath Routing for Multipartite State Distribution in Quantum Networks}
\author{\uppercase{Evan~Sutcliffe}\authorrefmark{1}, \IEEEmembership{Member, IEEE}, and
\uppercase{Alejandra~Beghelli\authorrefmark{1,2}} \IEEEmembership{Member, IEEE}}
\address[1]{Electronic \& Electrical Engineering Department, University College London, London, WC1E 7JE, UK}
\address[2]{BT Group, Ipswich, Suffolk, IP5 3RE, UK}
 
\tfootnote{This work was supported by the Engineering and Physical Sciences Research Council [grant number EP/S021582/1], TRANSNET (EP/R035342/1) and Innovate UK 10004793.}

\markboth
{Author \headeretal: Preparation of Papers for IEEE Transactions on Quantum Engineering}
{Author \headeretal: Preparation of Papers for IEEE Transactions on Quantum Engineering}

\corresp{Corresponding author: Evan Sutcliffe (evan.sutcliffe.20@ucl.ac.uk).}

\begin{abstract}
We consider the problem of distributing entangled multipartite states across a quantum network with improved distribution rate and fidelity. For this, we propose fidelity-aware multi-path routing protocols, assess their performance in terms of the rate and fidelity of the distributed Greenberger-Horne-Zeilinger (GHZ) states, and compare such performance against that of single-path routing. 
Simulation results show that the proposed multi-path routing protocols select routes that require more Bell states compared to single-path routing, but also require fewer rounds of Bell state generation. We also optimised the trade-off between distribution rate and fidelity by selecting an appropriate cutoff to the quantum memory storage time. Using such a cutoff technique, the proposed multi-path protocols can achieve up to an 8.3 times higher distribution rate and up to a 28\% improvement in GHZ state fidelity compared to single-path routing.
These results show that multi-path routing both improves the distribution rates and enhances fidelity for multipartite state distribution.
\end{abstract}

\maketitle

\IEEEpeerreviewmaketitle

\section{Introduction}
Quantum networks, made of devices able to exchange quantum information over long distances, are expected to enable many applications in quantum information processing \cite{cirac1999distributed,wehner2018quantum,cuomo2020towards,main2025distributed}. However, qubits should not be transmitted directly over long distances, as any encoded quantum information can be readily lost to interactions with the environment. Encoding quantum information in logically protected states can prevent this \cite{muralidharan2016optimal}, but such techniques are not feasible using current technologies \cite{Preskill2018nisq}. Instead, the reliable transmission of quantum information can be achieved using shared entangled states as a resource. For example, a maximally entangled two-qubit state, such as a shared Bell state, can be used to transmit the state of a single qubit via quantum teleportation \cite{bennett1996Distill}. 

To distribute Bell states, also described as bipartite entanglement, between non-adjacent nodes, quantum repeaters can be used \cite{azuma2021tools,Briegel1998repeater,munro2015inside,goodenough2021optimizing}. 
For two devices that each share Bell states with a repeater node, entanglement swapping can be performed on these two Bell states to produce a single Bell state shared between these non-adjacent devices. To share a Bell state between two distant devices over a quantum network, entanglement swapping can be performed recursively along a path of quantum repeaters between the devices.

While sharing quantum information between two devices requires access to a bipartite state, communication among multiple parties requires a shared \textit{multipartite} state. Multipartite states, such as the Greenberger-Horne-Zeilinger (GHZ) and graph states, are multi-qubit entangled states with many potential uses in quantum information processing \cite{hein2004multiparty,Pirker2019stack,Xu2014GHZQKD,oslovich2024efficientpublished,zhang2021distributed,Singh2024modulararchitecturesentanglementschemes,nickerson2013topological,deBone2024thresholds,Diadamo2021alphavqedist}. The GHZ state in particular can be used for multi-party quantum key distribution (QKD) \cite{Xu2014GHZQKD,oslovich2024efficientpublished}, quantum sensing \cite{zhang2021distributed}, quantum error correction \cite{Singh2024modulararchitecturesentanglementschemes,nickerson2013topological,deBone2024thresholds} and distributed quantum computing \cite{yimsiriwattana2004GHZCOMP,Diadamo2021alphavqedist,dhondt2004computational}.

To distribute a GHZ state among non-adjacent users, routing is a key task that must be performed. Routing is the selection of a set of devices and channels that are used to distribute quantum states between the users at the least cost. We call this set of devices and channels the \textit{routing solution}. 
Routing approaches can be classified into single-path and multi-path routing. Single-path routing involves the selection of a unique pre-computed routing solution. In contrast, in multi-path routing, the routing solution can be selected dynamically, from potentially many possible options \cite{pant2019routing,sutcliffetqe}. When multi-path routing is used, the network performs Bell state generation over many redundant channels, with the routing solution selected depending on the probabilistic outcomes of the Bell state generations.

Routing can be optimised by selecting different metrics. For quantum networks with noisy operations, both the distribution rate and fidelity of the distributed states should be considered \cite{munro2015inside,li2021efficient,inesta2023optimal,caleffi2017optimal,bugalho2023distributing}, especially as many applications, such as quantum error correction, have strict thresholds for the minimum fidelity of distributed states\cite{nickerson2013topological,deBone2024thresholds,sutcliffe2025distributed}. Previous research has shown that using multi-path routing can increase distribution rates \cite{pant2019routing,sutcliffetqe,patil2020GHZFuse,patil2021Multiplexed,zhao2021redundant,oslovich2024efficientpublished,Shi2024qcast}. However, so far the impact of multi-path routing on the fidelity of the distributed states has not been considered.

We investigate, for the first time, the problem of distributing multipartite GHZ states using fidelity-aware multi-path routing. We propose two multi-path protocols aimed at distributing $N$-qubit GHZ states to an arbitrary set of nodes in a noisy quantum network. For this, both the rate and fidelity of the distributed states are considered when selecting the routing solution. We also apply a cutoff time to quantum memories to control the minimum fidelity of Bell states over the network.
The distribution rate and fidelity achieved by the protocols are evaluated and compared against equivalent single-path routing protocols.

The contributions of this work are:
\begin{itemize}
    \item Two novel multipartite state distribution protocols that make use of fidelity-aware multi-path routing. The protocols use knowledge of the fidelity and location of Bell states over the network to perform routing dynamically. 
    
    \item The assessment, for the first time, of the relationship between multi-path routing and the fidelity of the distributed multipartite states. Further, we demonstrate regimes in which multi-path protocols distribute GHZ states at both a higher rate and fidelity than single-path routing.
    
    \item The assessment of the impact of a memory cutoff time when using multi-path routing. By selecting an appropriate cutoff time, we show how the trade-off between rate and fidelity can be optimised.
\end{itemize}

The rest of the paper is outlined as follows. In Section \ref{section:background}, we review the necessary background literature. In Section \ref{sec:model}, the network model is described, along with the qubit error model and the effect of this on the distributed states fidelity. In Section \ref{sec:proposal}, the new protocols are presented. The results and concluding remarks are presented in Sections \ref{sec:result} and \ref{sec:conc} respectively. \\

\vspace{1cm}

\section{Background} \label{section:background}

\subsection{Routing approaches and impact on distribution rate and fidelity}

\begin{figure*}[h!tb]
    \centering \includegraphics[width=0.9\linewidth]{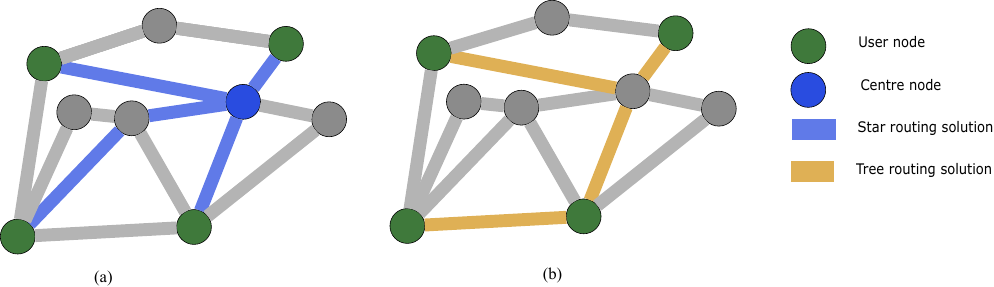}
    \caption{An example quantum network with quantum devices (nodes) connected by quantum channels (edges). Routing is calculated for distributing a GHZ state between the example users (green nodes). (a) An example of single-path routing using the star (sp-s) approach. The routing solution consists of an edge-disjoint path between each user to a centre node. (b) An example of the single-path tree approach (sp-t) in which uses the Steiner tree. Bell state generation is attempted over the edges in the routing solution until a GHZ state can be distributed.}
    \label{fig:treestar}
\end{figure*}

For single-path routing, the routing solution is a unique precomputed set of nodes and channels that are used to distribute a desired state. Bell state generation is attempted over the selected routing solution and then used to distribute the required entangled state. Given the probabilistic generation of non-local Bell states, the distribution rates will decrease with the distance between the users and with the number of Bell states required \cite{pant2019routing,azuma2021tools}. For multi-path routing, the routing solution is selected dynamically, using knowledge of the location of shared Bell states held over the network. Multi-path routing has been shown to improve the distribution rate of bipartite and multipartite states with respect to single-path routing \cite{acin2007entanglement,pant2019routing,patil2020GHZFuse,patil2021Multiplexed, sutcliffetqe,Shi2024qcast}, even when measurements are imperfect \cite{patil2020GHZFuse,van2023entanglement}. For bipartite state distribution, work to improve the fidelity of the distributed states has considered an iterative approach to the allocation of multiple paths. For this, entanglement distillation is performed recursively on the Bell state distributed over each path \cite{leone2024costvectoranalysis}. However, each path operates independently and these works do not consider a dynamic routing problem. 

For multipartite state distribution, the use of multi-path routing has generally focused on improving the distribution rate. 
One of the key reasons for this is that multi-path routing can achieve distance-independent rates, when percolation is observed in a network. This is not possible using single-path routing, where the distribution rate decreases with distance \mbox{\cite{azuma2021tools,pant2019routing,acin2007entanglement,sutcliffetqe}}. In probabilistic networks, where edges are independently present or absent with a fixed probability, long-distance connectivity is enabled by bond-percolation behaviour. When the edges (bonds) are present with a probability above the \textit{percolation threshold} $p_c$, the likelihood of the user nodes being found in the same connected component becomes independent of the distance between them. For the square lattice topologies considered, the percolation threshold is given by $p_c = \frac{1}{2}$. The reason for this long distance connectivity is that above the percolation threshold, a single giant connected component emerges that spans the majority of the network nodes, with all other connected components being small \mbox{\cite{grimmett2013percolation,bollobas1984evolution}}. For multi-path routing that can use any set of edges, successful GHZ state distribution simply requires that all user nodes be found within the same connected component, where edges represented Bell states shared between two nodes. However, previous works on multi-path routing have not considered the impact on fidelity.

In quantum networks, routing can further be combined with sub-routines that improve the fidelity of the distributed states. Examples include entanglement distillation \cite{deutsch1996quantum,Jansen2022bilocal,krastanov2021hetroPure,DeBone2020GHZStates,wallnofer2019multipartite} and memory cutoffs \cite{li2021efficient,inesta2023optimal,Vardoyan2021switch1,Rozp_dek_2018,azuma2021tools}.
Such sub-routines can improve the fidelity of distributed states, but at the cost of reducing the achievable rate \cite{deutsch1996quantum,inesta2023optimal}. In the case of entanglement distillation, higher fidelity states are generated from multiple noisy copies of the state. However, the generation of multiple copies can significantly increase the total wait time required to distribute the entangled states.
Quantum memory cutoffs are limits to the maximum permitted storage time of a qubit in a quantum memory. Cutoffs can be implemented to address the reduction in fidelity that occurs due to decoherence. Decoherence describes the loss of quantum information over time when qubits are stored in noisy quantum memories. By using knowledge of the decoherence rate, a cutoff can be selected to discard a state when its fidelity falls below a predetermined threshold. 
The choice of cutoff can be adjusted to modify the rate and fidelity trade-off. For single-path routing, optimal cutoff times can be calculated analytically \cite{li2021efficient,inesta2023optimal}. However, such work has not been extended to multi-path routing, which requires more complex analytical expressions. 
\subsection{Multipartite state distribution} \label{sec:MSD}

Due to their use in quantum information applications, various works have considered the distribution of multipartite states in quantum networks \cite{bugalho2023distributing,avis2023analysis,Vardoyan2021switch1,Cacciapuoti2024Stop,DeBone2020GHZStates,pirker2018modular,wallnofer2019multipartite,meignant2019distributing}. A key theme of many of these works is the use of a central node in the routing solution. A central node can be used to generate a $N$-qubit multipartite state locally, with a qubit of the state then teleported out to each of the $N$ users \cite{avis2023analysis,pirker2018modular,Fischer2021StarGraphState}.
When distributing GHZ states, the approach can be simplified using \textit{entanglement fusion} \cite{DeBone2020GHZStates,patil2020GHZFuse}. In this case, each user must first generate a shared Bell state with the centre node. If a user is not adjacent to the centre node, this can be achieved via entanglement swapping along a path of devices. Then, by performing entanglement fusion between the qubits of the Bell states stored in the central node, the Bell states can be fused into a GHZ state, shared between the users \cite{avis2023analysis}. In both cases, the routing solution requires a separate path between each user and the central node. We term such routing approaches \textit{star routing}, with an example given in Fig.~\ref{fig:treestar}(a).
Further, the generation of a GHZ state by entanglement fusion does not require that all fusion operations be performed within a single centre node. In Meignant \textit{et al.} it was shown how a GHZ state can be generated from a tree of Bell states that connect the users \cite{meignant2019distributing}, where fusion is performed at nodes storing the qubits of multiple Bell states. The tree that connects a set of users at minimum cost is the Steiner tree \cite{pcst}. Fig.~\ref{fig:treestar}(b) shows a routing example using the Steiner tree, which we term \textit{tree routing} \cite{bauml2020linear,oslovich2024efficientpublished,sutcliffetqe}. 

The combined generation and distillation of GHZ states from entanglement fusion has been considered more generally, in cases for which routing is not required. Such work has considered situations where devices are adjacent or arbitrarily connected. Applications of such work include distributed quantum error correction, where high fidelity GHZ states must be distributed between fixed sets of four adjacent devices \cite{DeBone2020GHZStates,nickerson2013topological,deBone2024thresholds}. While these schemes do not consider how to distribute states long distances over quantum networks, techniques from such work could be applied to sharing high fidelity multipartite states between arbitrary devices in complex topologies.

\subsection{Routing for multipartite state distribution}

Single-path routing for multipartite state distribution has been studied considering both star \cite{bugalho2023distributing,avis2023analysis,Vardoyan2021switch1,Fischer2021StarGraphState} and tree approaches \cite{bugalho2023distributing,meignant2019distributing,sutcliffetqe,bauml2020linear}.
In Bugalho \textit{et al.}, star routing approaches were proposed \cite{bugalho2023distributing}, with Avis \textit{et al.} extending these approaches with implementation strategies as well as further analytical bounds on the distribution rate and fidelity \cite{avis2023analysis}. In Avis \textit{et al.} multiple rounds of Bell state generation are assumed to be required. 
For approaches that use a star approach, memory cutoffs have been considered as an approach to adjust the rate and fidelity trade-off. Such cutoffs have been optimised using Monte Carlo or Markov chain analysis \cite{Vardoyan2021switch1,Cacciapuoti2024Stop,kumar2023optimal}. Single-path routing to select a routing solution for a tree approach has been applied to GHZ state distribution \cite{bugalho2023distributing,bauml2020linear}. 

Multi-path routing has similarly been investigated for multipartite state distribution, considering both the star and tree approaches. It has been shown that multi-path routing can improve the distribution rate for sharing GHZ states over a quantum network \cite{sutcliffetqe}. Further work has extended this result to show improved multi-party QKD key rate \cite{oslovich2024efficientpublished}. However, the impact of multi-path routing on the fidelity of the distributed states, has not been considered. In noisy networks in which multiple rounds of Bell state generation are required, the relationship between decoherence and the higher distribution rate achieved by multi-path routing has also not been explored.

\section{Network and Error Models} \label{sec:model}

We consider protocols which distribute $N$-qubit multipartite GHZ states, defined as \hbox{$\ket{\text{GHZ}_N} = \frac{1}{\sqrt{2}} (\ket{0}^N +\ket{1}^N)$} over a quantum network. The performance of such multipartite distribution protocols will be sensitive to network parameters, such as the rate and fidelity at which Bell states can be generated. In the following, we define a quantum network model used for protocol performance evaluation.

\subsection{Network model}

\begin{figure}[h!tb]
    \centering \includegraphics[width=0.9\linewidth]{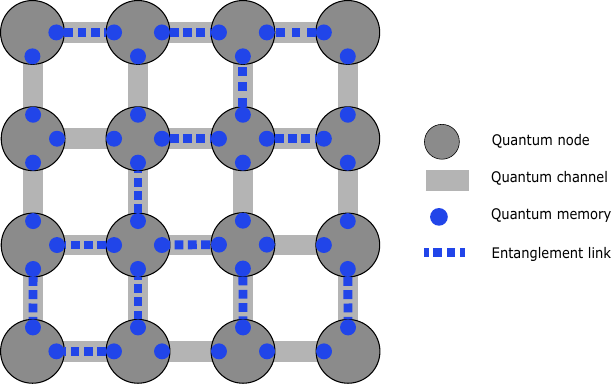}
    \caption{An example quantum network in a $4\times 4$ grid topology. The network consists of devices (nodes), that are interconnected by quantum channels (edges). Bell states (entanglement links) can be generated between adjacent nodes over a shared channel. Also shown is an example of generated entanglement links, that can be described using the entanglement link-state graph $G^\prime$.}
    \label{fig:graph}
\end{figure}
The quantum network is modelled as a graph \mbox{$G = (V,E)$,} which is a collection of nodes $v \in V$ connected by a set of edges $e \in E$ such as seen in Fig.~\ref{fig:graph}.
Nodes are devices equipped with a single quantum memory allocated for communication, for each adjoining edge. The nodes can store qubits in quantum memories as well as perform single-qubit, two-qubit and measurement operations. We assume that all nodes have identical capabilities.
An edge represents a quantum channel which can be used to share a Bell state of the form \hbox{$\ket{\phi^+} = \frac{1}{\sqrt{2}} (\ket{00} +\ket{11})$} between two adjacent nodes.
Such shared Bell states will henceforth be referred to as \textit{entanglement links}. 
Entanglement links are generated probabilistically, with probability $p_e$ for network edge $e$. A heralded entanglement distribution scheme is assumed, such that a successful entanglement link generation is flagged by a classical signal \cite{saha2024HighFidelity}. Given the single memory per edge assumption, at most one entanglement link can be stored over an edge and entanglement link generation cannot be re-attempted while the memories are still occupied. We also define a \textit{link-state graph} $G^\prime=(V,E^\prime)$, which describes the location of entanglement links over the quantum network. In $G^\prime$, an edge $e = (u,v)$ will be present when nodes $u$ and $v$ share an entanglement link. This definition of the link-state graph means that $G^\prime$ will contain all nodes from $G$, even if some do not hold any entanglement links. In such a situation, such nodes would simply be disconnected in $G^\prime$.
Finally, we model the operation of the network as being performed in discrete timeslots of duration $T_{\text{slot}}$. A timeslot includes the generation of entanglement links, as well as any local operations and classical communication (LOCC) required to generate the GHZ state. We consider the implementation of a quantum memory cutoff. In a discrete timeslot network model, a cutoff $Q_c$ defines the maximum number of timeslots an entanglement link can be stored in a quantum memory before being discarded. As the network operates over multiple timeslots, we use the index $t \in \{1,2,\ldots\}$ to denote the $t^\text{th}$ timeslot. One timeslot, of period $T_\text{slot}$ in the discrete-time network model is sometimes referred to as a \textit{round}. In this context, $k$ rounds would refer to an ordered set of $k$ sequential timeslots. Finally, as the state of $G^\prime$ will vary between timeslots, as entanglement links are generated and discarded, we define a link-state graph $G^\prime_{t}$, which represents the state of $G^\prime$ during timeslot $t$.

\subsection{Error model}
We consider multipartite state distribution in a network of noisy quantum devices, where imperfect Bell state generation and decoherence are the dominant sources of error. LOCC operations are assumed to be performed instantaneously and ideally which means that any noise in the distributed GHZ state can be attributed to noise from the Bell states used for distribution. Examples of LOCC operations include entanglement swapping and entanglement fusion.
We quantify the impact of noise on the distributed states by their fidelity \cite{Nielsen2002Bible}. 
The fidelity $\mathcal{F}$ is a measure of the overlap between two quantum states with $ 0 \leq \mathcal{F} \leq 1$. For a mixed state $\rho$, such as a state generated under noisy conditions, the fidelity can be defined with respect to a desired pure state $\ket{\Psi}$ as:
\begin{equation}
    \mathcal{F}(\Psi,\rho) = \bra{\Psi}\rho\ket{\Psi}
\end{equation}

We model quantum noise using the depolarising noise channel \cite{Nielsen2002Bible}, which has been shown to be a good model for noise in quantum networks and can be considered a worst-case error channel \cite{helsen2023benchmarking}. 
For an $N$-qubit mixed state $\rho$ with dimension $d=2^N$, the $N$-qubit depolarising channel transforms the state by: 
\begin{equation} 
    \mathcal{D}(\rho) = p_n \, \rho + \frac{(1-p_n)}{2^N} \mathbb{I}_d 
\end{equation}\label{eq:depol}
This has the action of leaving the state unchanged with probability $p_n$ and transforming the state into the maximally mixed state $\mathbb{I}_d/d$ with probability $1-p_{n}$. Here $\mathbb{I}_d$ is the $d$-dimension identity matrix and the maximally mixed state $\mathbb{I}_d/d$ represents a complete loss of information about the prior quantum state. For entanglement links distributed under noisy conditions, the effect of depolarising noise can be described using Werner states \cite{bennett1996Distill}:
\begin{equation}
    \rho_w = w \, \ket{\phi^+}\bra{\phi^+} + \frac{1-w}{4} \mathbb{I}_4
\end{equation}
A Werner state is a mixed state which is composed of the ideal Bell state with probability $w$, and the maximally mixed state with probability $1-w$.
The fidelity of a Werner state with respect to the desired Bell state can be given in terms of the Werner parameter $w$ as $\mathcal{F} =\frac{3 w +1}{4}$ \cite{bennett1996Distill}.

As well as modelling noise during entanglement link generation, the effect of storing the entanglement links in noisy quantum memories can also be quantified using Werner states \cite{li2021efficient,inesta2023optimal}.
A Werner state stored for time $\tilde{t}$ in two imperfect memories will have a Werner parameter given by \hbox{$w_{\tilde{t}} = w_0 e^{-\tilde{t}/T_c}$}, where $T_c$ is a measure of decoherence time for the Bell state and $w_0$ is the initial value of $w$, due to noisy entanglement link generation, at time $\tilde{t}=0$. For a discrete-timeslot network model, this expression can be rewritten using the age of the Werner state $\tau$ and a grouped decoherence constant $\Delta = e^{- T_{\text{slot}}/T_c}$: 
\begin{equation} 
    w_\tau = w_0 \Delta^{\tau} \label{eq:2}
\end{equation}
The decoherence constant $\Delta$ also represents the probability of both qubits in the Werner state being unaffected by single-qubit depolarisation noise (e.g., $ 1-\Delta = (1-p_n)^2$) during the time period $T_{\text{slot}}$. As such, $\Delta$ will be a function of both the decoherence rate and the timeslot length.
The age of a Werner state $\tau \in \{0,1,\ldots\}$ is defined as the number of timeslots after generation that the state has been stored in memory. It is computed as the difference between the index of the current timeslot and that of the timeslot in which the state was generated.
For a given minimum Bell state fidelity, an appropriate cutoff $Q_c$ can be calculated using Eq.~(\ref{eq:2}) such that states are discarded once they fall below a threshold fidelity. However, a smaller value of the cutoff $Q_c$ will also reduce the end-to-end distribution rate \cite{azuma2021tools}. 

\subsection{GHZ state generation and fidelity} \label{sec:gen}

The distribution of a multipartite GHZ state over a quantum network first requires a set of Bell states (entanglement links) to be generated between the users. Then, these Bell states can be combined into the desired GHZ state using LOCC. The fidelity of the distributed state will therefore be a function of the fidelity of the Bell states consumed, as well as the exact configuration of how swapping and fusion operations are applied. While we assume that these LOCC operations are executed ideally, non-ideal LOCC operations can also be modelled by considering additional noise as being applied to the Bell states \cite{bugalho2023distributing}, or simulated numerically.

\begin{figure}
    \centering
    \includegraphics[width=0.7\linewidth]{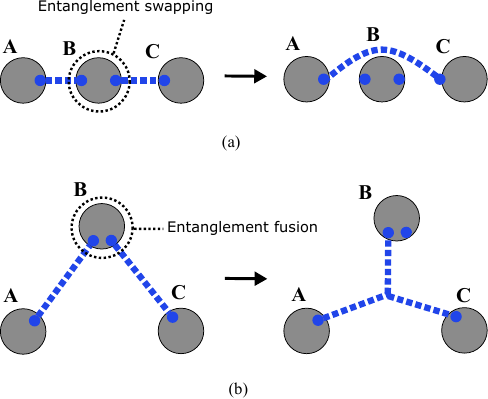}
    \caption{(a) An intermediate repeater node B performs a Bell state measurement (BSM) to produce a Bell state shared by nodes A and C. (b) Node B performs entanglement fusion on two Bell states to produce a 3-qubit GHZ state shared between nodes A, B and C.}
    \label{fig:swapping}
\end{figure}

Entanglement swapping is an operation that can be used to generate a shared Bell state between two nodes not connected by a quantum channel. Fig.~\ref{fig:swapping}(a) shows how entanglement swapping can be performed by a quantum repeater. If the two end nodes A and C both share a Bell state with an intermediate repeater node B, then by performing a Bell state measurement (BSM) on the qubits held at the repeater node, a Bell state shared between nodes A and C is generated. By performing entanglement swapping recursively along a path of quantum repeaters, this approach can be used to distribute a Bell state between two arbitrary nodes in a quantum network \cite{azuma2021tools,munro2015inside}. 

If entanglement swapping is performed using noisy Bell states, the state distributed to the end nodes will also be affected by noise. For entanglement swapping performed on Werner states, $w_1$ and $w_2$, the state outputted will also be a Werner state with $w_{out}=w_1 \cdot w_2$. For Werner states, the effect of depolarisation noise can be combined multiplicatively when entanglement swapping is performed \cite{li2021efficient}.
This relationship can be extended to entanglement swapping performed along a path of quantum repeaters. The state distributed between the end nodes will have a Werner parameter:
\begin{equation} \label{eq:prod_b}
     w_{B}=\prod_{e \in B} w_e
\end{equation}
where $w_e$ represents the Werner parameter held over an edge $e$ and $B$ represents a path, which could be a branch in a routing solution. In a star routing solution, a branch is a path between the centre node and each user node. In a tree routing solution, the branches are the set of paths between the user and \textit{fork} nodes. A fork node is a node in the Steiner tree with a nodal degree $\geq 3$. A branch will not contain user or fork nodes, except at its endpoints. The fidelity of the Bell state distributed by entanglement swapping over a branch is given by $\mathcal{F}_B = (3 w_{B} + 1)/4$. 

Entanglement fusion can be used to combine two GHZ states of $n_1$ and $n_2$ qubits respectively, into an \hbox{($n_1+n_2-1$)-qubit} GHZ state \cite{nickerson2013topological,DeBone2020GHZStates,chelluri2024multipartite}. Entanglement fusion is an example of a LOCC operation, so requires at least one qubit of each GHZ state to be held in the same node.

Fig.~\ref{fig:swapping}(b) shows how a 3-qubit GHZ state can be generated by fusing two Bell states\footnote{Note that the $\ket{\phi^+}$ Bell state is equivalent to a $2$-qubit GHZ state.}.
The fidelity of a GHZ state generated by entanglement fusion between two noisy GHZ states will depend on the noise and the choice of qubits that are used for the fusion operation.
The fidelity of a GHZ state, $\mathcal{F}_{\text{GHZ}}$, distributed using the star approach, can be expressed as a function of the fidelity of the Bell states $\mathcal{F}_B$ to be fused (one per branch $B$) \cite{bugalho2023distributing}: 

\begin{equation}
\begin{split}
\mathcal{F}_{\text{GHZ}}  =  & \frac{1}{2} \left( \prod_{B \in R} \frac{4 \mathcal{F}_B-1}{3} + \prod_{B \in R} \frac{2(1-\mathcal{F}_B)}{3} \right.  \\
&  \left.  + \prod_{B \in R} \frac{1+ 2\mathcal{F}_B}{3} \right)
\end{split}
\end{equation}
This expression assumes that the state to be distributed is between $|S|>2$ users, so that there are multiple branches in $R$.
Similarly, for GHZ states generated from a tree of Bell states, the GHZ state fidelity also has an analytical description in which there are recursive terms dependent on the fidelity and arrangement of Bell states in the tree
\cite{bugalho2023distributing}.

A final operation for GHZ state distribution is qubit removal. A qubit can be removed from a GHZ state, without destroying the entanglement between the remaining qubits, by performing an $X$-basis measurement on the qubit to be removed. Assuming this operation is executed ideally, there is no impact on the fidelity of the remaining GHZ state.

\section{Fidelity-aware multipartite distribution protocols} \label{sec:proposal}

\begin{figure*}[h!t]
    \centering
    \includegraphics[width=0.85\linewidth]{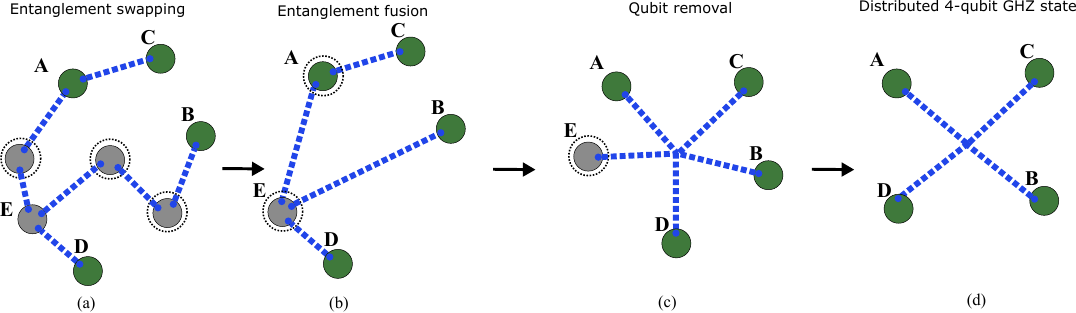}
    \caption{Generation of a 4-qubit GHZ state shared between the four users (green nodes A-D) from a tree of 7 Bell states by swapping, fusion and qubit removal. At each step, the corresponding operations are performed in the nodes circled with a dashed line.}
    \label{fig:fuse_tree}
\end{figure*}

In this section, we propose multi-path routing protocols for distributing multipartite GHZ states in a fidelity-aware manner. The protocols aim to distribute a GHZ state between a set of \textit{users} $S$ which are a subset of nodes $S \subseteq V$, from a network graph $G$ as described in Section \ref{sec:model}.
We assume the users wish to share a single $N$-qubit GHZ state, such that each $s \in S$ obtains a qubit of the GHZ state ($|S| = N$).
First, we describe a generic multipartite distribution protocol in terms of three distinct phases: routing, entanglement link generation, and GHZ state generation, to then introduce the specifics of the proposed protocols.

\subsection{Multipartite state routing} \label{sec:routing}
We consider multipartite state distribution protocols which use either the star or tree approaches introduced in Section \ref{section:background}. For each, we define a routing solution $R$ as a subgraph of $G$ in which every edge must hold an entanglement link before a GHZ state can be distributed ($ R\subseteq G^\prime$). For the star routing approach, $R$ consists of a set of paths (branches) between a central node $v_c$ and each user $s \in S$ that are edge-disjoint in $G^\prime$.
For the tree routing approach, $R$ can be any tree that connects the users. A formal description of the star and tree routing problems are given in Appendix \mbox{\ref{app:formal}}.

A main objective when selecting $R$ is to maximise the distribution rate by selecting edges with the highest probability of successfully generating all the required entanglement links across $R$ \mbox{\cite{azuma2021tools}}. This can be achieved by selecting a routing solution $R$ that maximises $p_{R}$:

\begin{equation} 
    p_{R} = \prod_{e \in R}{p_e} \label{eq:prod_p} 
\end{equation} 

A routing solution that maximises $p_R$ (or any multiplicative cost function) can be found by using the logarithm of the edge parameters as a weight (e.g. $c_e=-\log(p_e)$) with standard routing algorithms such as Dijkstra's shortest path \mbox{\cite{dijkstra1959note}}, minimum-cost max-flow \mbox{\cite{orlin1997polynomial}} or the Steiner tree algorithm \mbox{\cite{pcst}}.

Selecting a routing solution that maximises GHZ state fidelity is more complex. The exact GHZ fidelity will depend nonlinearly on the fidelity of each Bell state over each branch in $R$. This means selecting $R$ by directly optimising for the GHZ state fidelity, in particular for the tree routing approaches, this makes direct optimisation computationally infeasible. To calculate the Steiner tree, standard approaches to routing which assume linear-cost edge-weights are known to be \hbox{NP-Hard}, so any optimisation of fidelity directly would be at least as hard computationally.
We therefore make use of a fidelity lower bound $\mathcal{F}_{LB}$ ($ \mathcal{F}_{GHZ} \geq \mathcal{F}_{LB}$ ) as a proxy that can be efficiently optimised:

\begin{equation} \label{eq:fi}
\mathcal{F}_{LB} =  \prod_{e \in R} \mathcal{F}_{e}
\end{equation}

The lower bound follows from the fact that fidelity $\mathcal{F}_e$ of a Werner state over edge $e$ is also equal to the probability that it is found in the desired $\ket{\phi^+}$ state \mbox{\cite{bennett1996Distill}}. The value of $\mathcal{F}_{LB}$ therefore denotes the probability that all Werner states in $R$ are found in the $\ket{\phi^+}$ state, which would result in the generation of a GHZ state generated with fidelity at least $\mathcal{F}_{LB}$ \cite{wallnofer2019multipartite}.
We also consider the term $w_R$, which is the product of Werner parameters from $R$ as: 
\begin{equation}
     w_{R} = \prod_{e \in R} w_e  \label{eq:prod_r}
\end{equation}
For Werner states the relationship between $\mathcal{F}$ and $w$ is given by $\mathcal{F} = (3w+1)/4$. As the transformation between $\mathcal{F}$ and $w$ is affine (linear up to a constant) and monotonic, using either $w_e$ or $\mathcal{F}_e$ for routing will select the same optimal routing solution $R$ that maximises $\mathcal{F}_{LB}$ assuming $w \in [0,1]$. As such, $\mathcal{F}_{LB}$ can be maximised using standard routing techniques with edge-weights \mbox{$c_e = -\log(w_e)$}. As $\mathcal{F}_e \geq w_e$, we can further say that $\mathcal{F}_{LB} \geq w_R$. 
Further discussion on the motivation and accuracy of the fidelity lower bounds are given in Appendix \mbox{\ref{app:f_lower_bound}}, in which we show that the difference between the lower bound and exact GHZ state fidelity was found to be, on average less than $1.1\%$.

\subsection{Entanglement link generation} 
As entanglement link generation is probabilistic, it can be reattempted for multiple rounds until all required Bell states have been generated. This process is performed over a selected subset of edges in the graph selected for multipartite state distribution. After an entanglement link has been generated, it is stored in noisy quantum memories during which time it undergoes decoherence modelled using its Werner parameter $w_e$ as described in Eq. (\ref{eq:2}). 
As a cutoff (of $Q_c$ timeslots) is utilised, after an entanglement link has been stored for $\tau_e \geq Q_c$ timeslots, it is discarded. Given the assumption of each node having a single quantum memory allocated for communication per network edge, once an entanglement link is generated over an edge, entanglement link generation cannot be reattempted over that edge until it has been consumed or discarded.

\subsection{GHZ state generation} \label{sec:m_gen_scheme}

We describe an approach for generating a GHZ state from a tree or star of Bell states shared over a quantum network. However, other equivalent schemes could also be used \cite{meignant2019distributing,avis2023analysis}. Once the routing solution has been selected, the means of distributing the GHZ state is identical for both the single-path and multi-path routing approaches. The scheme proposed assumes a tree layout but, as all star topologies are also trees, the scheme is valid for both the star and tree approaches. Fig.~\ref{fig:fuse_tree} shows how the operations described can be used to transform a tree of entanglement links into a GHZ state shared between the users.
The tree routing solution $R$ is first reduced by performing entanglement swapping along any branches in the tree.
This results in a set of Bell states shared between the nodes at the end-points of each branch. These nodes will be the users nodes, as well as any fork nodes of the routing solution.
Entanglement fusion is then performed at all nodes in this reduced tree that hold multiple qubits of the Bell states. 
After the entanglement fusion operations, a GHZ state has been generated that is shared between the users and fork nodes. A final step is to remove any qubits from the GHZ state held at the fork nodes by means of an $X$-basis measurement. The resulting state, up to locally applied corrections, is an $N$-qubit GHZ state shared between the set of users $S$, such that each user holds a single qubit of the GHZ state. Qubit removal is an optional operation that can be performed to ensure the distributed state is exactly an $N$-qubit GHZ state shared between the user nodes $S$. Otherwise, the state distributed would be a larger GHZ state with qubits also held at the `fork' nodes. For some applications, a larger GHZ state will not be an issue but for others, such as QKD, this would not be acceptable.

\subsection{Baseline and proposed protocols} \label{sec:protocols}
We describe four multipartite entanglement distribution protocols. Two are baseline protocols used for benchmarking that utilise the single-path star (sp-s) and tree (sp-t) approaches. Two are novel protocols that use fidelity-aware multi-path routing. A star (mp-s) and tree (mp-t) variant of the proposed multi-path routing approach are described. 
The single-path star protocol is adapted from the work described in \cite{bugalho2023distributing,avis2023analysis} for distributing GHZ states in noisy quantum networks. Similarly, the single-path tree protocol is a generalisation of single-path approaches which consider a Steiner tree for routing \cite{bugalho2023distributing,sutcliffetqe,meignant2019distributing,bauml2020linear}. {The benchmark single-path protocols are designed to operate under identical conditions as the proposed multi-path protocols excluding routing. While there are more advanced techniques that can be combined with routing for distributing entangled states, this allows for a direct comparison of the routing approaches. For the multi-path protocols (mp-s and mp-t), the protocols require the link-state graph $G^\prime$ to be known for routing. Obtaining the state of $G^\prime$ at a central controller will require additional classical communication. Any latency due to this additional classical communication will have an impact on the fidelity of the states distributed due to decoherence. However, the latency requirements should not be significant for certain qubit technologies and moderate-sized quantum networks, such as quantum data-centres. The impact of latency on entanglement distribution protocols is an important area of research but is out of the scope of this work.

\subsubsection{Benchmark single-path protocols}

The single-path star (sp-s) and single-path tree (sp-t) protocols are the benchmark single-path protocols that use the star and tree routing approaches respectively. The routing solution $R$ of the sp-s protocol consists of a set of paths from a centre node $v_c$ to each user in $s \in S$. These paths must be edge-disjoint in the link-state graph $G^\prime$. 
For the sp-t protocols, the routing solution is a Steiner tree that connects $S$. Example routing solutions for sp-s and sp-t are shown in Fig.~\mbox{\ref{fig:treestar}}. A formal definition of the routing approaches used by the star and tree protocols is given in Appendix \mbox{\ref{app:formal}}. For benchmarking we simulate the protocols in networks $G$ with uniform edge parameters ($p_e,w_0,\Delta,Q_c$). This ensures that the single-path routing solution is optimal for both distribution rate and fidelity, as any routing solution that maximises rate will also maximise fidelity.

\begin{figure*}
    \centering
    \includegraphics[width=0.85\linewidth]{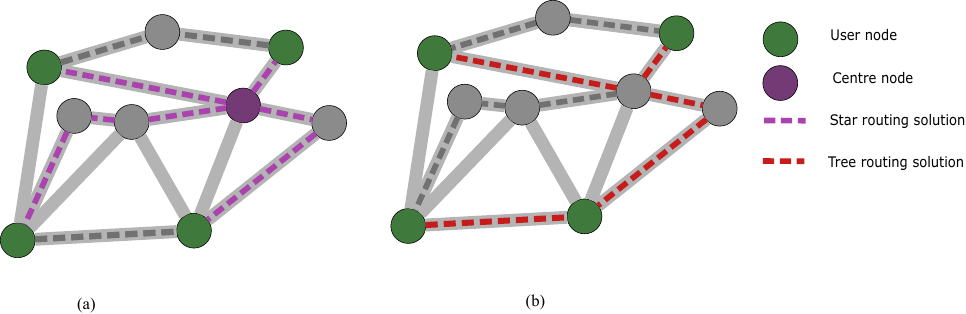}
    \caption{Multi-path star and tree routing solutions for distributing a GHZ state. For the multi-path approaches, routing is attempted with knowledge of the global link-state graph $G^\prime$ (dashed lines). In this example, (a) shows multi-path routing using a star approach (mp-s), which selects a routing solution of 7 edges, (b) shows the multi-path tree approach (mp-t), where the routing solution requires 5 edges.}
    \label{fig:network_mps_mpt}
\end{figure*}

\begin{algorithm}[h!tb]
\caption{Single-path routing (star/tree)}\label{alg:singlepath}
\small
\begin{algorithmic}
\Function{SinglePathRouting}{$G, S, \text{type}$}
    \State $G^\prime_{0} \gets (V, \varnothing)$
    \State $(R, v_c) \gets$ GetRoutingSolution($G, S, \text{type}$)
    \State $t \gets 0$
    \Repeat
        \State $t \gets t + 1$
        \State $G^\prime_{t} \gets$ DiscardCutoff($G^\prime_{t-1}$)
        \State $G^\prime_{t} \gets$ EntanglementLinkGeneration($R, G^\prime_{t}$)
        \If{$R \subseteq G^\prime_{t}$}
            \State GHZ $\gets$ GenerateGHZFromRoutingSolution($G^\prime_{t}, R, S$)
            \State \Return GHZ
        \EndIf
    \Until{GHZ state successfully generated}
\EndFunction
\end{algorithmic}
\end{algorithm}

The protocols are described together in pseudocode form in Algorithm~\mbox{\ref{alg:singlepath}}. The two protocols have similar operations, and so are described together with the routing condition differentiated by the variable $\text{type} \in \{\text{star},\text{tree}\}$. Initially, the routing solution $R$ is selected, as denoted by the function \texttt{GetRoutingSolution}, from $G$ to maximise the distribution rate and fidelity. The function \texttt{GetRoutingSolution} represents the routing algorithms formally described in Appendix~\mbox{\ref{app:formal}}, to select the optimal single-path star or tree routing solution. The sp-s protocol also requires the selection of a centre node $v_c$. As all nodes in the network are assumed to be equally equipped, any node can operate as the centre node. As such, the optimal routing solution $R$ is calculated for each node $v \in V$ operating as centre node. Of these candidate solutions, the $(R,v_c)$ pair which maximises the rate and fidelity is selected.

Once $R$ has been selected, the temporal evolution of the network begins. The link-state graph $G^{\prime}$ in timeslot $t$ is described as $G^{\prime}_t$. At the start of each timeslot, entanglement links with an age older than the cutoff are discarded (\texttt{DiscardCutoff}) and then entanglement link generation (\texttt{EntanglementLinkGeneration}) is attempted. These functions model the operation of the network nodes, and are described further in Algorithm~\mbox{\ref{alg:discard}} and \mbox{\ref{alg:generate}}. The cutoff $Q_c$ is defined as an edge-parameter over $G$, but in Section~\mbox{\ref{sec:result}} we assume it to be uniform. Entanglement link generation is then attempted over each edge $e \in R$ of the routing solution that does not already hold an entanglement link, with generation success sampled, by the entanglement link generation probability $p_e$, described as the function \texttt{GenerateLinkSuccess}. If all edges in $R$ hold an entanglement link ($R \subseteq G^{\prime}_{t}$) then a GHZ state is generated (\texttt{GenerateGHZFromRoutingSolution}). This function represents the operations described in Section~\mbox{\ref{sec:m_gen_scheme}}, to generate a shared GHZ state from a set of Bell states. If a GHZ cannot be generated, the protocol proceeds to the next timeslot until a GHZ state has been generated. While there is no implicit guarantee on successful distribution within a specific time window, the protocols are assumed to continue until successful GHZ state distribution.

When evaluating the protocols via Monte Carlo simulation, an additional parameter $t_\text{max}$ was introduced to bound the simulation runtime. If a GHZ state is not successfully generated within $t_\text{max}$ timeslots, the run is marked as a failure. The value of $t_\text{max}$ was chosen to be large compared to the expected number of timeslots to ensure accurate convergence of the distribution rate. This is a feature of the sampling approach, rather than the protocols, and for this reason is excluded from the pseudocode. Further, for this reason, $t_\text{max}$ is not intended to be a parameter to be adjusted to modify the behaviour of the protocol.

\begin{algorithm}[h!tb]
\caption{Discard entanglement links using a fixed cutoff}
\label{alg:discard}
\small
\begin{algorithmic}
\Function{DiscardCutoff}{$G^{\prime}$}
\ForAll{$e \in G^{\prime}$}
    \If{$\tau_e \geq Q_c(e)$}
        \State Remove $e$ from $G^{\prime}$
    \EndIf
\EndFor
\State \Return $G^{\prime}$
\EndFunction
\end{algorithmic}
\end{algorithm}

\begin{algorithm}[h!tb]

\caption{Entanglement link generation where $R \subseteq G$ can represent a routing solution (single-path) or the whole graph $G$ (multi-path) depending on the protocol.}\label{alg:generate}
\small
\begin{algorithmic}
\Function{EntanglementLinkGeneration}{$R, G^{\prime}$}
\ForAll {$e \in R$}
    \If{$ e \notin G^{\prime}$ \textbf{and}
        GenerateLinkSuccess$(e)$}
            \State Add $e$ to $G^{\prime}$
    \EndIf
\EndFor
\State \Return $G^{\prime}$
\EndFunction
\end{algorithmic}
\end{algorithm}

\subsubsection{Multi-path protocols}
The proposed fidelity-aware multi-path protocols are the multi-path star (mp-s) and multi-path tree (mp-t) protocol. Example routing solutions for the mp-s and mp-t protocols are shown Fig.~\mbox{\ref{fig:network_mps_mpt}}. Here, the routing solutions are selected from $G^\prime$. This approach will not necessarily give routing solutions that are optimal compared to if routing was performed in $G$. However, if a valid routing solution exists, it can be used immediately.

The operation of the protocols are described in Algorithm~\mbox{\ref{alg:multipath}} where the star or tree routing condition is signified by the variable \textit{type}.
For the multi-path protocols, the only task required before temporal evolution is the selection of a centre node $v_c$ for the mp-s protocol. The centre node $v_c$ is selected using the same approach as the sp-s protocol but if multiple candidate centre nodes give the same cost, $v_c$ is chosen to be the centroid of $S$. We assume the centre node is fixed, however it could also be selected dynamically.

In temporal evolution, first the memory cutoff is applied (\texttt{DiscardCutoff}), and then entanglement generation (\texttt{EntanglementLinkGeneration}) is attempted. For multi-path routing, entanglement link generation is attempted over all edges in $G$ that do not already hold an entanglement link. Then, \texttt{RoutingFeasible} is checked to assess if it is possible to distribute a GHZ state. For the mp-s protocol, a set of edge-disjoint paths will exist between $v_c$ and each $s \in S$ as long as the minimum cut between $v_c$ and $S$ in $G^\prime$ exceeds $|S|$ \mbox{\cite{orlin1997polynomial}}. For the mp-t protocol, a Steiner tree that spans $S$ can be found as long as all users $s \in S$ are in the same connected component of $G^\prime$. If routing is feasible, then $R$ is selected from $G^\prime_{t}$ using \texttt{GetMultipathRoutingSolution} to maximise the GHZ state fidelity. The function \texttt{GetMultipathRoutingSolution} represents the routing algorithms described in Appendix~\mbox{\ref{app:formal}}, applied to the multi-path routing problem. That is, the optimal star or tree routing solution are selected from the link-state graph $G^\prime$. By performing routing over $G^\prime$, the routing solution can be selected to maximise the GHZ state fidelity, using knowledge of the location and fidelity of the entanglement links in $G^\prime$, instead of having to consider the joint optimisation of the distribution rate and fidelity when routing is performed offline.
GHZ states are distributed from the entanglement link of $R$ using the method of generating a GHZ state from Bell states described in Section~\mbox{\ref{sec:m_gen_scheme}}.  
If routing is not feasible, the protocol proceeds to the next timeslot.

\begin{algorithm}[h!tb]
\caption{Multi-path routing (star/tree)}\label{alg:multipath}
\small
\begin{algorithmic}
\Function{MultiPathRouting}{$G, S, \text{type}$}
    \State $G^\prime_{0} \gets (V, \varnothing)$
    \State $v_c \gets$ None
    \If{$\text{type} == \text{star}$}
        \State $(\_, v_c) \gets$ GetRoutingSolution($G, S, \text{type}$)
    \EndIf
    \State $t \gets 0$
    \Repeat
        \State $t \gets t + 1$
        \State $G^\prime_{t} \gets$ DiscardCutoff($G^\prime_{t-1}$)
        \State $G^\prime_{t} \gets$ EntanglementLinkGeneration($G, G^\prime_{t}$)
        \If{RoutingFeasible($G^\prime_{t}, S, \text{type}$)}
            \State $R \gets$ GetMultipathRoutingSolution($G^\prime_{t}, S, v_c, \text{type}$)
            \State GHZ $\gets$ GenerateGHZFromRoutingSolution($G^\prime_{t}, R, S$)
            \State \Return GHZ
        \EndIf
    \Until{GHZ state successfully generated}
\EndFunction
\end{algorithmic}
\end{algorithm}

We initially also considered multi-path routing protocols which made use of n-qubit fusion operations (GHZ-basis projective measurements) between the qubits of all Bell states held at each node in the network \mbox{\cite{patil2020GHZFuse,clayton2024efficient}}. Such protocols operate by fusing all entanglement links in a network, rather than just along a selected routing solution. The benefit of such approaches is that they can be executed using only local link-state information. However, we found such approaches achieved a worse distribution rate and fidelity compared to the proposed multi-path protocols when Bell states were distributed in the presence of depolarising noise. The relative performance of such approaches are discussed in Appendix \mbox{\ref{app:B}}.

\subsection{Computational complexity}

\begin{table}[ht!] 
    \caption{Computational complexity of offline and online routing algorithms}
    \label{tab:comp1}
    \centering
         \begin{tabular}{|p{1cm}|p{3cm}|p{3cm}|}
     \hline
    \textbf{Protocol}&\textbf{Offline computational complexity }&\textbf{Online computational complexity}\\
     \hline
    sp-s & $O(|V|^2|E|^2)$ & $O(|R|)$ \\ \hline
    sp-t & $O(|E| \text{log} |V|)$  & $O(|R|)$ \\ \hline
    mp-s & $O(|V|^2|E|^2)$ & $O(|V||E|^2)$ \\ \hline
    mp-t & $O(1)$ & $O(|E| \text{log} |V|)$  \\ \hline 
    \end{tabular}
\end{table}

\begin{table*}[h!tb] 
    \caption{Network parameter default values. The first four parameters define the quantum network, while the last two describe the protocol configuration.}
    \label{tab:comp}
    \centering
         \begin{tabular}{|c|c|p{10cm}|}
     \hline
    \textbf{Parameter}& \textbf{Default Value} & \textbf{Description}\\
    \hline
    \hline
    $G$ & $6\times 6$ grid & Network graph $G= (V,E)$: Graph describing the network topology, Default graph is a $6 \times 6$ square grid with  $|E|=60$, $|V|=36$. \\ 
    $p$ & $0.1$ & Entanglement link generation probability: the probability of generating a Bell state between two adjacent nodes over an edge.\\
    $w$ & 0.987 & Werner parameter: defines the probability that entanglement links are generated without depolarising noise. \\ 
    $\Delta$ & 0.99 & Decoherence constant:  probability that entanglement links are unaffected by depolarisation when stored in quantum memories.\\
    \hline \hline
    $S$ & $|S|=4$ & Users: Set of nodes ($S \subseteq V$) requiring a shared $N$-qubit GHZ state. Each set of users $S$ was randomly selected from $V$.\\
    $Q_c$ & $[1,20]$ & Memory cutoff time: the maximum number of timeslots an entanglement link is stored before being discarded. \\ \hline
    \end{tabular}
\end{table*}

The computational complexity of the routing protocols are described in Table \mbox{\ref{tab:comp1}}. For this, we separate tasks that are performed offline, i.e. before entanglement link generation begins and tasks performed online i.e. \mbox{per timeslot}. The offline tasks performed by the single-path protocols are the calculation of the pre-determined routing solutions $R$. For the sp-s protocol, the routing solution is calculated using a minimum-cost max-flow algorithm (with complexity\footnote{The complexity term can also be described as $O(|V||E|^2CU)$ where $C:=\text{max}(f_e)$ where $f_e$ is the edge flow capacity, and $U$ is the maximum sum nodal capacity. For the grid topologies considered, these terms are constant $C=1$ and $U=4$ and so are excluded. For more general topologies and edge capacities these terms may become relevant.} $O(|V||E|^2)$) repeated up to \mbox{$|V|$} times, once for each candidate centre node \mbox{$v_c \in V$} \mbox{\cite{király2012efficientimplementationsminimumcostflow}}.
For the sp-t protocol the routing solution can be calculated using a Steiner tree routing algorithm. While this is an NP-hard problem, it can be solved with complexity $O(|E| \text{log} |V|)$
using approximate approaches with bounded accuracy \mbox{\cite{pcst}}. For both single-path protocols, the online (per timeslot) complexity is $O(|R|)$, as the only operation required is checking that all edges in $R$ hold an entanglement link ($R \in G^\prime$).

The multi-path approaches have a reduced requirement for offline tasks.
The mp-s protocol requires the offline selection of a centre node, which is performed in the same manner and hence with the same complexity, as the sp-s protocol. The mp-t protocol has constant time offline tasks which have complexity $O(1)$. In comparison to the single-path approaches, the multi-path routing protocols require more complex online (per timeslot) operations. For the mp-s protocol, the online task required is to calculate the star routing solution using the max-flow approach with complexity $O(|V||E|^2)$,  \mbox{\cite{király2012efficientimplementationsminimumcostflow}}. Similarly the mp-t protocol has a per timeslot computational complexity of solving the approximate Steiner tree problem.

The described routing algorithms compute and utilise the fidelity lower bound $\mathcal{F}_{LB}$ when selecting the optimal routing solution to maximise the GHZ state fidelity. If required, the fidelity lower bound for an arbitrary routing solution $R$ can be calculated with linear complexity complexity $O(|R|)$.

\section{Results} \label{sec:result}

This section presents the performance evaluation of the protocols proposed in Section~\ref{sec:proposal}. The protocols were assessed in terms of the GHZ state distribution rate and fidelity. Results were generated\footnote{The code used to generate the results is available at \hbox{\url{https://github.com/evansutcliffe/qmcs}}.} by means of Monte Carlo simulations. 

\subsection{Performance metrics and model parameters}
The distribution rate is the rate at which GHZ states are shared between the users. For a discrete-time network model, the distribution rate ($\lambda$) can be defined as the reciprocal of the expected number of rounds $\mathbb{E}[T]$ (each round being of length $T_{\text{slot}}$) of entanglement link generation required to distribute a GHZ state:
\begin{equation} 
    \lambda = \frac{1}{\mathbb{E}[T]}
\end{equation}
The second metric considered is the fidelity, which quantifies the impact of noise on the distributed states. For a noisy GHZ state described by the mixed state $\rho$, the fidelity $\mathcal{F}_\text{GHZ}$ can be defined with respect to the desired GHZ state \cite{Nielsen2002Bible}:
\begin{equation}
    \mathcal{F}_\text{GHZ} =\bra{\text{GHZ}_N} \rho \ket{\text{GHZ}_N}
\end{equation} 

In the simulations, the fidelity was calculated by implementing the approach described in Section~\ref{sec:m_gen_scheme} using NetSquid \cite{Coopmans2021netsquid}. These values were also validated against analytical expressions \cite{bugalho2023distributing}.

Simulations were conducted on a $6\times 6$ square grid topology with randomly selected subsets \hbox{$S \subseteq V $} of $|S| =4$ nodes requiring shared $4$-qubit GHZ states. We consider only grid topologies, however the protocols described are applicable to any network topology.
Each simulation run was independent, with multiple rounds of entanglement link generation (timeslots) being attempted until a GHZ state was distributed and the protocol terminated.
Each data point was repeated and averaged over 100 randomly selected sets of users $S$. For each, the simulation was run until $300$ GHZ states had been distributed, or for up to $3\times 10^6$ total timeslots, whichever occurred first. This was achieved by setting $t_\text{max}=10^4$ for each run. While $t_\text{max}$ was chosen to be large relative to the expected number of timeslots required to distribute a GHZ state, the distribution statistics will depend on the location of the users and underlying network topology. A data point is not shown where either one of the 100 sets of $S$ achieved zero successful GHZ state distributions or fewer than $200$ total successful GHZ states were distributed.

Table \ref{tab:comp} outlines the default parameters used in the simulations. When non-default values are used, they are specified with the result. The entanglement link generation probability $p$ is the probability of an entanglement link being generated during a timeslot which is assumed to be uniform for all edges ($p_e = p, \;\forall e \in E$). We assume a default value of $p = 0.1$, a value that is suggested to be feasible with near-term quantum devices \cite{Singh2024modulararchitecturesentanglementschemes}.
The initial Werner parameter $w_0$ and decoherence constant $\Delta$ describe the impact of noise on the entanglement links in the network. The values of $w_0=0.987$ and $\Delta=0.99$ are motivated by experimental results by Saha \textit{et al.} in which Bell states were generated between trapped ions qubits in separate devices \cite{saha2024HighFidelity}. Bell state fidelities of $\mathcal{F} = 0.97$ were achieved, with claims $\mathcal{F} > 0.999$ will be achievable with improved hardware. Motivated by these results, we assume Bell states are generated with a fidelity of $\mathcal{F}=0.99$ ($w_0=0.987$). 
The value of $\Delta =0.99$ was selected based on experimental Bell state generation systems, to reflect typical decoherence rates for a near-term quantum network \cite{saha2024HighFidelity,main2025distributed}. We model the implementation of a quantum memory cutoff, $Q_c$, that is uniform over all quantum memories in the network. We considered cutoffs in the range $Q_c \in [1,20]$, where a cutoff of $Q_c = 1$ represents entanglement links being discarded at the end of every timeslot if a GHZ state cannot be distributed.

\subsection{Distribution rate and fidelity trade-off} \label{sec:res1}

Fig.~\ref{fig:jointDR_F} shows both the distribution rate and fidelity achieved by the protocols.
Each data point shows the average rate and fidelity achieved for over 100 configurations of user sets $S \subseteq V$, where $|S|=4$ and each $S$ was randomly selected from $V$. 
For each protocol the simulations were repeated for a set of cutoff times $Q_c \in [1,20]$, with the value of $Q_c$ annotated on the data point. 
Sweeping $Q_c$ shows how increasing the cutoff leads to GHZ states being distributed at a higher rate, but with a reduced fidelity. The Pareto frontier can be observed, which shows that the mp-t protocol had the best performance in the simulated network. 
Comparing the best performing multi-path protocol (mp-t) to the best single-path protocol (sp-t), multi-path routing gave an improvement in distribution rate of up to $8.3$ times and an improvement in GHZ state fidelity of up to $28\%$. The improvement in distribution rate (resp. fidelity) was calculated by comparing the maximum improvement in distribution rate between the mp-t and sp-t protocols, such that the other parameter, e.g. fidelity (resp. rate), of mp-t protocol is equal to or better than the sp-t protocol. The maximum distribution rate speedup was achieved when comparing the mp-t protocol to the sp-t protocol operating with $Q_c = 13$, and the fidelity improvement when compared to the sp-t protocol with $Q_c = 20$.

When comparing the star approaches (mp-s and sp-s), we see the mp-s protocol achieved an improvement in distribution rate of up to $\times 2.2$ and $16\%$ improvement in fidelity. 
Further, we see that for both single-path and multi-path routing, the tree approaches (sp-t and mp-t) outperformed the star approaches (sp-s and mp-s). As the tree approaches select the least cost routing solution, the tree approaches, by definition, select routing solutions that give a higher distribution rate and fidelity. This is because the star routing solutions are valid, but not necessarily optimal, tree solutions. Results for different entanglement link generation probabilities $p \in \{0.2,0.3\}$, found in Appendix \ref{app:res_app}, show the same overall trends.

We focus on topologies in the form of regular grid lattices. However, the protocols described can operate over any network topology including random topologies such as Erd\H{o}s-R\'enyi graphs. The multi-path routing protocols considered will perform better in network topologies with many alternative paths and when the average path length between users is low.

 \begin{figure}[h!tb]
    \centering
    \includegraphics[width=\linewidth]{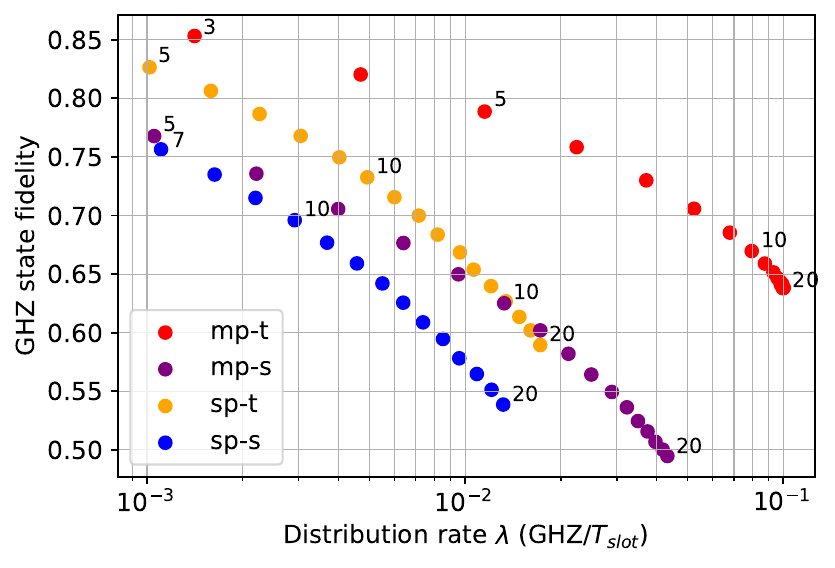}
    \caption{The distribution rate and fidelity achieved by the mp-t, mp-s, sp-t, and sp-s protocols for $Q_c \in [1,20]$ (annotated). Parameters: $ G = 6\times 6$ grid, $p = 0.1$.}
    \label{fig:jointDR_F}
\end{figure}
\subsection{Routing solution size and entanglement link age}

The fidelity of the GHZ states distributed depends on the number of entanglement links required and their fidelities. To quantify this, we consider how the impact of depolarising noise over the routing solution $w_R$ can be separated in terms of the noise during Bell state generation, and due to decoherence. Assuming that $w_0$ and $\Delta$ are uniform across the network the fidelity lower bound from Eq. (\ref{eq:prod_r}) can be combined with Eq. (\ref{eq:2}):
\begin{equation}
    \mathcal{F}_{\text{GHZ}} \geq \prod_{e \in R} w_e = \prod_{e \in R} w_0 \Delta^{\tau_e} 
     = w_0^{|R|} \Delta^ {\bar{\tau} |R|}
\end{equation}
This shows how the noise applied to the entanglement links can be described in terms of two main components. The first component describes the impact of noise that occurs during Bell state generation, and is dependent on the size $|R|$ of the routing solution.
The second component describes the impact of decoherence and depends on both the average age $\bar{\tau}$ and the number of required entanglement links $|R|$. Here, age is defined as the number of timeslots an entanglement link has been stored in a quantum memory when it is used for generating a GHZ state. 

\begin{figure}[h!tb]
    \centering
    \subfloat[\centering]
    {{\includegraphics[width = \linewidth]{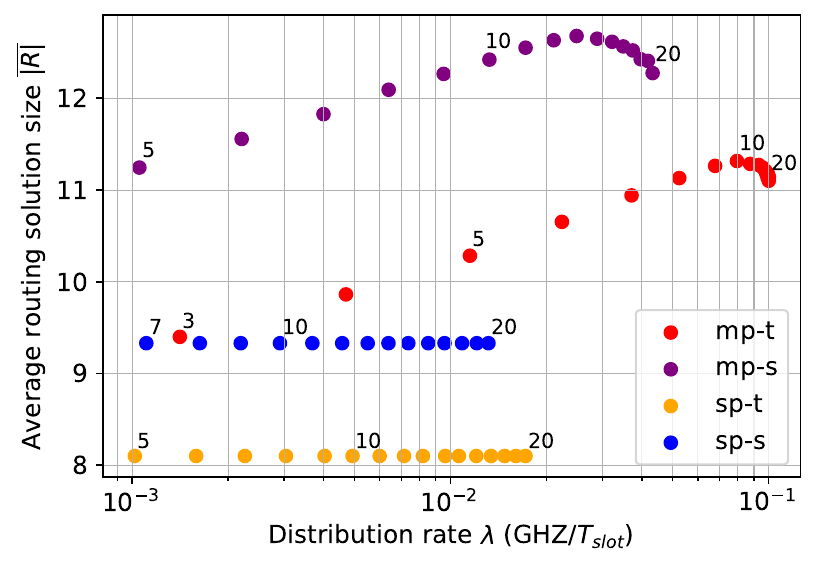}\label{fig:r_size}}}
    \qquad
    \subfloat[\centering]
    {{\includegraphics[width =\linewidth]{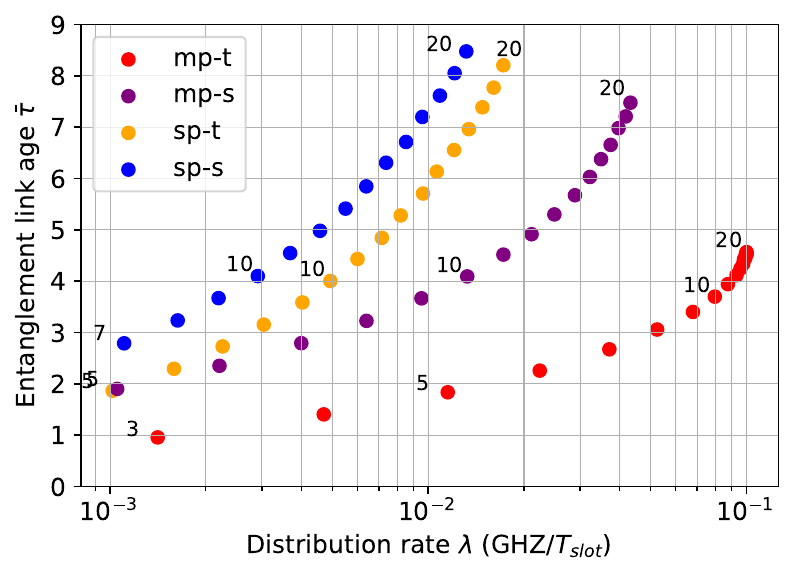}\label{fig:tau_bar}}}
    \qquad
    \caption{(a) Average routing solution size $\overline{|R|}$ against the distribution rate, where $\overline{|R|}$ is defined as the average number of entanglement links used for GHZ state generation. (b) The average age {$\bar{\tau}$} in terms of number of timeslots, of the entanglement links in $R$ when they are utilised for GHZ state generation. The protocols were simulated for $Q_c \in [1,20]$ (annotated). Parameters: $G = 6\times 6$ grid, $p = 0.2$, $Q_c \in [1,20]$.}
    \label{fig:r_size+tau_bar}
\end{figure}

Figures \ref{fig:r_size+tau_bar}(a) and \ref{fig:r_size+tau_bar}(b) show the routing solution size and entanglement link age against distribution rate, for the same simulation scenario as Fig.~\ref{fig:jointDR_F}.
In single-path routing the routing solution is pre-computed such that $R$ and hence $|R|$ are fixed for all values of $Q_c$.
In contrast, the results show how multi-path protocols select routing solutions with a size that increases with $Q_c$ and the achieved distribution rate. While the distribution rate is strictly increasing with $Q_c$, the routing solution size $|R|$ observes a peak for some value of $Q_c$. The multi-path protocols make use of routing solutions which are up to 36\% larger for star approaches and 40\% larger when the tree approach is used. Using larger routing solutions reduces the fidelity of the states distributed due to the $w_0^{|R|}$ term. Fig.~\ref{fig:r_size+tau_bar}(b) shows the distribution rate against the average age, $\bar{\tau}$, of the entanglement links on the $y$-axis. For all protocols, increasing the cutoff $Q_c$ increases the average age of entanglement links used. In Fig.~\ref{fig:r_size+tau_bar}(b), the results show that the multi-path protocols select routing solutions in which the entanglement links have a lower age and a much higher distribution rate. Due to the lower average age of the entanglement links used by the multi-path routing protocols, the impact of noise due to decoherence is reduced. Further, for the multi-path protocols, increasing $Q_c$ leads to smaller increases in age compared to the single-path protocols. This shows how multi-path routing can improve the fidelity of the distributed states by mitigating the impact of decoherence.

\subsection{Distance dependent performance}

\begin{figure}[h!tb]
    \centering
    \includegraphics[width=\linewidth]{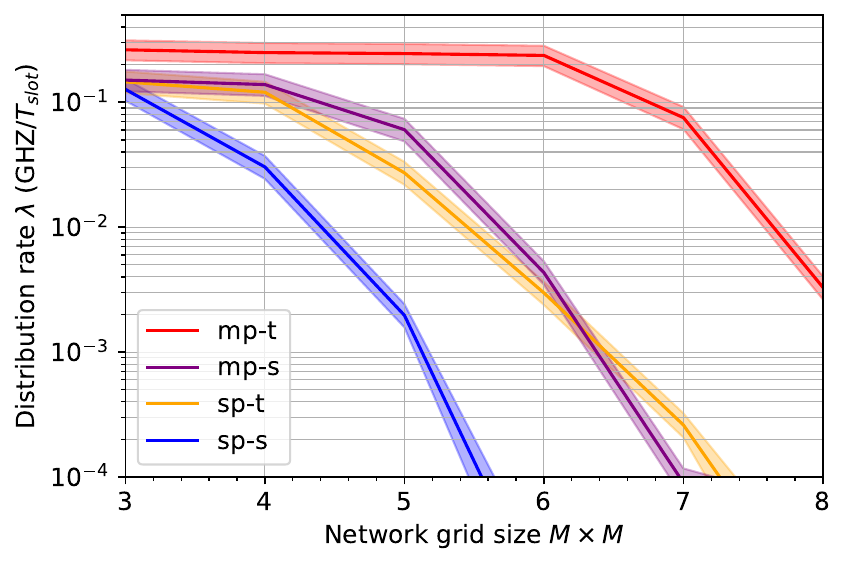}
    \caption{Distribution rate for different protocols with users in corner nodes of $M \times M$ grid topology. For each distance, the value of $Q_c$ was selected to maximise distribution rate while maintaining an average GHZ fidelity $\mathcal{F}_\text{GHZ}> 2/3$. Parameters: $p=0.3$, $Q_c \leq 20 $.}
    \label{fig:distance}
\end{figure}

Multi-path routing has been shown to achieve a distribution rate that is independent of the distance between the users \cite{pant2019routing,patil2020GHZFuse,sutcliffetqe}. However, these works assume noiseless or error-corrected Bell states and so these results do not generally hold for noisy quantum networks. For a state distributed over a quantum network using noisy entanglement links with $w_e<1$ $\forall e \in R$, the fidelity of the distributed state will be strictly decreasing with $|R|$.
We define the distance between multiple users in the network in terms of the Steiner distance $D_G(S)$, which is the size of the Steiner tree that connects the users in $S$.
Therefore, for noisy Bell states, the GHZ states distributed will have a fidelity that is strictly decreasing with distance irrespective of the routing approach used.

In Section~\ref{sec:res1} it was shown how the cutoff $Q_c$ can be adjusted to optimise the trade-off between rate and fidelity. Using this approach, we assess the rate at which the protocols can distribute GHZ states above a minimum fidelity $\mathcal{F_\text{GHZ}} \geq 2/3$ with increasing distance. The protocols were simulated for distributing a 4-qubit GHZ state between users located in the four corner nodes of an $M \times M$ grid network with $p=0.3$. By increasing the size of the $M \times M$ grid topology, the Steiner distance between the users increased by $D_G(S) =3(M-1)$. For the protocols, the value of $Q_c$ was optimised for each distance to maximise the distribution rate while maintaining an average fidelity $\mathcal{F_\text{GHZ}} \geq 2/3$. Fig.~\ref{fig:distance} shows the distribution rate against network size $M$ for the network model considered. Error bars show the minimum likelihood estimation of the distribution rate with a 99.9\% confidence interval as implemented by \cite{Gidney2021stim}. 
As entanglement link generation is probabilistic, increasing the distance between the users leads to a lower distribution rate when single-path routing is used \cite{azuma2021tools}. Further, as the required size of the routing solution increases with distance, lower values of $Q_c$ must be selected to maintain the minimum GHZ fidelity, which further reduces the distribution rate \cite{praxmeyer2013reposition}. 

For multi-path routing, a distance-independent distribution rate can be achieved if entanglement links are generated with a uniform probability $p$ above the percolation threshold $p_c$. For graphs with edges generated (and discarded) over multiple timeslots, percolation can be achieved over the link-state graph $G^{\prime}_{t}$ as long as:
\begin{equation} \label{eq:x0}
    P[e \in E_t'] > p_c
\end{equation}

The probability $P[e \in E_t']$ of an entanglement link being present in $E'$ at timeslot $t$, is equal to the probability of one successful generation within the previous $k$ rounds assuming $k\leq Q_c$.
\begin{equation} \label{eq:x1}
    P[e \in E_t'] = 1-(1-p)^k
\end{equation}

From Eq. (\mbox{\ref{eq:x0}}) and Eq. (\mbox{\ref{eq:x1}}) we can find the minimum value of $k$ required to exceed the percolation threshold of a given topology:

\begin{equation} \label{eq:x2}
    k \geq \frac{\log(1-p_c)}{\log(1-p)}
\end{equation}

For a value of $Q_c$ such that $Q_c \geq k$ with $k$ from Eq. \mbox{(\ref{eq:x2})}, multi-path routing exhibits a distance-independent behaviour. Where a cutoff value of $Q_c \geq k$ is selected, a constant distribution rate will be achieved with $\mathbb{E}[T] \approx k$ and $\lambda \approx 1/k$. This is observed for the multi-path tree protocol for up to the $M= 6$ grid (distance $D_G(S) = 15$). 
However, with increasing distance between the users (with increasing $M$), a smaller cutoff $Q_c$ is required to achieve the minimum fidelity requirement. As the distance $D_G(S)$ increases, there becomes a point for which it is infeasible to distribute a GHZ with a fidelity above the minimum acceptable fidelity while maintaining $Q_c \geq k$. Beyond this point, the distribution rate decreases exponentially with increased distance. 

For the specific set of users located in the corner nodes, we see that a $79$ times improvement in the distribution rate was achieved in the $6 \times 6$ grid topology using multi-path routing. This is the worst case distribution rate in the network for single-path routing and hence the improvement due to multi-path routing is larger. In the average case, the routing solutions required to connect the users are smaller, so GHZ states are distributed with a higher rate and fidelity. Due to the additional requirements for a valid star routing solution, e.g. all users being connected to the centre node by an edge-disjoint path, we do not see distance independent behaviour for the mp-s protocol for the network considered. For less stringent network conditions, such as larger $p,w,\Delta$, distance independent behaviour can be observed.

\section{Conclusions and further work} \label{sec:conc} 
We have proposed protocols for distributing multipartite states which use multi-path routing, to select a route dynamically, using knowledge of the location and fidelity of Bell states over the network. 
The performance of such protocols was compared to single-path approach with fixed routing, considering both the rate and fidelity at which GHZ states are distributed. 
Results show that the multi-path protocols distributed GHZ states at a  higher rate and fidelity compared to single-path routing. 
The higher fidelity was achieved due to the reduced impact of noise due to decoherence when using multi-path routing. 
Finally, by selecting an appropriate quantum memory cutoff time, we showed up to an $8.3\times$ increase in distribution rate and up to a 28\% improvement in fidelity compared to the single-path routing approach.
For the worst-case set of users, the distribution rate achieved by multi-path routing was up to 78 times higher.
These results suggest that multi-path routing could be used to improve performance on near-term quantum networks, particularly when qubit operations are noisy.

Possible areas for further work includes modelling the impact and scalability due to classical latency. For multi-path routing there are potential trade-offs between the improved performance observed and an increased impact of decoherence due to higher classical latency. Another area of further work is the evaluation of the performance of multi-path routing in topologies, and for demands, that are likely to be utilised in near-term quantum networks, specifically considering the scalability of such approaches. Another area of further work could be integrating multi-path routing with more advanced entanglement distribution techniques, for example, swap-when-ready approaches or non-uniform memory cutoffs \mbox{\cite{li2021efficient,goodenough2025noise}}. 
For example, the integration of entanglement distillation with multi-path routing. Multi-path routing could be used to generate multiple multipartite states for entanglement distillation \mbox{\cite{leone2024costvectoranalysis}} or perform entanglement distillation on multiple entanglement links held over an edge. Probabilistic entanglement distillation schemes can have a significant impact on distribution rate but the higher tolerance of multi-path protocols to probabilistic edge failures may be especially beneficial.

\bibliographystyle{IEEEtran}
\bibliography{IEEEabrv,bib}

\appendices
\section{Formal Problem Statement}
\label{app:formal}
The optimisation problem considered consists of finding a routing solution $R$ that connects the users $S$ in a manner that maximises the GHZ state fidelity. In place of the exact fidelity, we maximise the fidelity lower bound $\mathcal{F}_{LB}$, which is the product of edge fidelities $\mathcal{F}_e$ where $\mathcal{F}_e = (3w_e+1)/4$. We focus on routing to maximise fidelity because the multi-path protocols perform routing on a fixed link-state graph $G^\prime$ such that only the entanglement link fidelities need to be considered. For the single-path protocols, we can also consider the problem as a fidelity only maximisation as for benchmarking purposes, we consider networks $G$ where edge parameters $p,w,Q_c,\Delta,$ are uniform. This ensures that the routing solution $R$ that maximises distribution fidelity will also be the routing solution that maximises the rate in the networks considered.

Routing for the tree approach consists of finding the Steiner tree $R$ that spans $S$ in $G^\prime$ to maximise fidelity. The problem of selecting the Steiner tree can be expressed as:
\begin{align*}
\text{minimise} \quad & \sum_{e \in R} c_e \\
\text{subject to} \quad & R \text{ is connected} \\
& S \subseteq R
\end{align*}

By choosing edge-weights $c_e = -\log(w_e)$, the Steiner tree selected will maximise $\mathcal{F}_{LB}$. For Bell states, the transformation between $\mathcal{F}$ and $w$ is affine and monotonic ($w \in [0,1]$), so using either $-\log(\mathcal{F}_e)$ or $-\log(w_e)$ as edge-costs for the routing algorithms considered will yield the same optimal routing solution.

For the star based protocols, routing is performed to select the least-cost set of edge-disjoint paths between a centre node $v_c$ and each user $ s \in S'$ where $S' = S \setminus \{v_c\}$, accounting for the case where a user also operates as the centre node $v_c$ and hence does not require an additional path. The optimal routing solution can be found by transforming the problem into a min-cost max-flow problem \mbox{\cite{király2012efficientimplementationsminimumcostflow}}. We define a flow $ f: E_f \to \{0,1\} $ over unit-capacity edges. The optimisation problem then becomes:
\begin{align*}
\text{minimise} \quad & \sum_{e \in E_f} c_e f_e \\
\text{subject to} \quad
& \sum_{(v_c, v) \in E_f} f_{v_c v} - \sum_{(v, v_c) \in E_f} f_{v v_c} = |S'| \\
& \sum_{(u, s) \in E_f} f_{u s} - \sum_{(s, u) \in E_f} f_{s u} = -1, \quad \forall s \in S' \\
& \sum_{(u, v) \in E_f} f_{u v} - \sum_{(v, u) \in E_f} f_{v u} = 0, \\
& \hspace{4em} \forall v \in V \setminus (S \cup \{v_c\}) \\
& f_e \in \{0,1\}, \quad \forall e \in E_f
\end{align*}
The routing solution $R$ can then be constructed from the edges of $E_f$ for which $f_{e} = 1$. The routing solution $R$ is accepted only if the total flow from $v_c$ equals $|S|$ and there is unit flow to each user. These conditions ensure that $R$ corresponds to a set of edge-disjoint paths from $v_c$ to each user $s \in S$. By minimising the cost of edge-weights $c_e = -\log(w_e)$ the fidelity term $\mathcal{F}_{LB}$ is maximised. For mp-s, the flow is calculated using edges from $G^\prime_{t}$ and the value of $w_e$ in timeslot $t$. For the single-path routing protocols, $w_e = w_0$ and routing is calculated considering all edges from $G$.

\section{Fidelity lower bound} 
\label{app:f_lower_bound}
A fidelity lower bound is used as a computationally feasible metric for selecting the least-cost routing solution. The fidelity of a multipartite state generated from fusing a set $K$ of noisy Bell states (of fidelity $\mathcal{F}_i$) can be lower bounded by \mbox{\cite{wallnofer2019multipartite}}:
\begin{equation}
    \mathcal{F}_{LB} = \prod_{i \in K} {\mathcal{F}_i}
\end{equation}
This expression assumes perfect LOCC operations and depolarising noise. This lower bound is useful for routing as it does not depend on the arrangement of which Bell states are being fused together. This can be seen using an equivalent (up to rearranging) description of a Werner state \cite{bennett1996Distill}:

\begin{equation}
\begin{split}
\rho_w =   \mathcal{F} &\ket{\phi^+}\bra{\phi^+} +  \\ & \frac{1 - \mathcal{F}}{3} \left( \ket{\phi^-}\bra{\phi^-} + \ket{\psi^+}\bra{\psi^+} + \ket{\psi^-}\bra{\psi^-} \right)
\end{split}
\end{equation}
\noindent where $\mathcal{F} = (3w+1)/4$ and $\{\ket{\phi^+},\ket{\phi^-},\ket{\psi^+},\ket{\psi^-}\}$ are the four Bell states \cite{bennett1996Distill}. In this description, the fidelity $\mathcal{F}$ also describes the probability that the Werner state is found in the desired $\ket{\phi^+}$ state. For a multipartite state generated from a set of Werner states $K$, the lower bound $\mathcal{F}_{LB}$ is the probability that all Werner states are found in the desired $\ket{\phi^+}$ state. 
The mixed state description of a multipartite state generated from such Werner states can be described as a combinatoric expansion of noise being applied to a subset of $K$ \mbox{\cite{avis2023analysis,tanizawa2023fidelityestimation}}. In such a description, higher-order terms describe the contribution to the mixed state of one or more Bell states being found in an undesired state. The fidelity contribution from these higher order terms will be strictly positive. As a result, a GHZ state generated using the routing solution $R$ will have an exact fidelity strictly higher than the product form lower bound $\mathcal{F}_{LB}$.

The term $\mathcal{F}_{LB} = \prod_{e \in R} \mathcal{F}_e$ describes a fidelity lower bound in terms of the fidelity of entanglement links in $R$.
However the accuracy of this bound can improved if we consider that most of the entanglement links are used to generate long-range Bell states via entanglement swapping, which has a simple analytical description for fidelity. The fidelity of a Bell state generated over a branch $B$ of the routing solution $R$ by entanglement swapping will be:
\begin{equation}\label{eq:fi_2}
    \mathcal{F}_B = \frac{3 \prod_{e \in B} w_e +1}{4}
\end{equation}
By using the exact fidelity of Bell states generated over the branches of $R$ ($B\subseteq R$), a tighter lower bound on the GHZ state fidelity $F_{LB'}$ can be calculated:

\begin{equation} \label{eq:fi_3}
    \mathcal{F}_{LB'} = \prod_{B \subseteq R} \mathcal{F}_{B}
\end{equation}
This tighter approximation is a post-processing operation that has no impact on the selection of the routing solution.
\begin{figure}[h!tb]
    \centering
    \subfloat[\centering]
    {{\includegraphics[width = 0.9\linewidth]{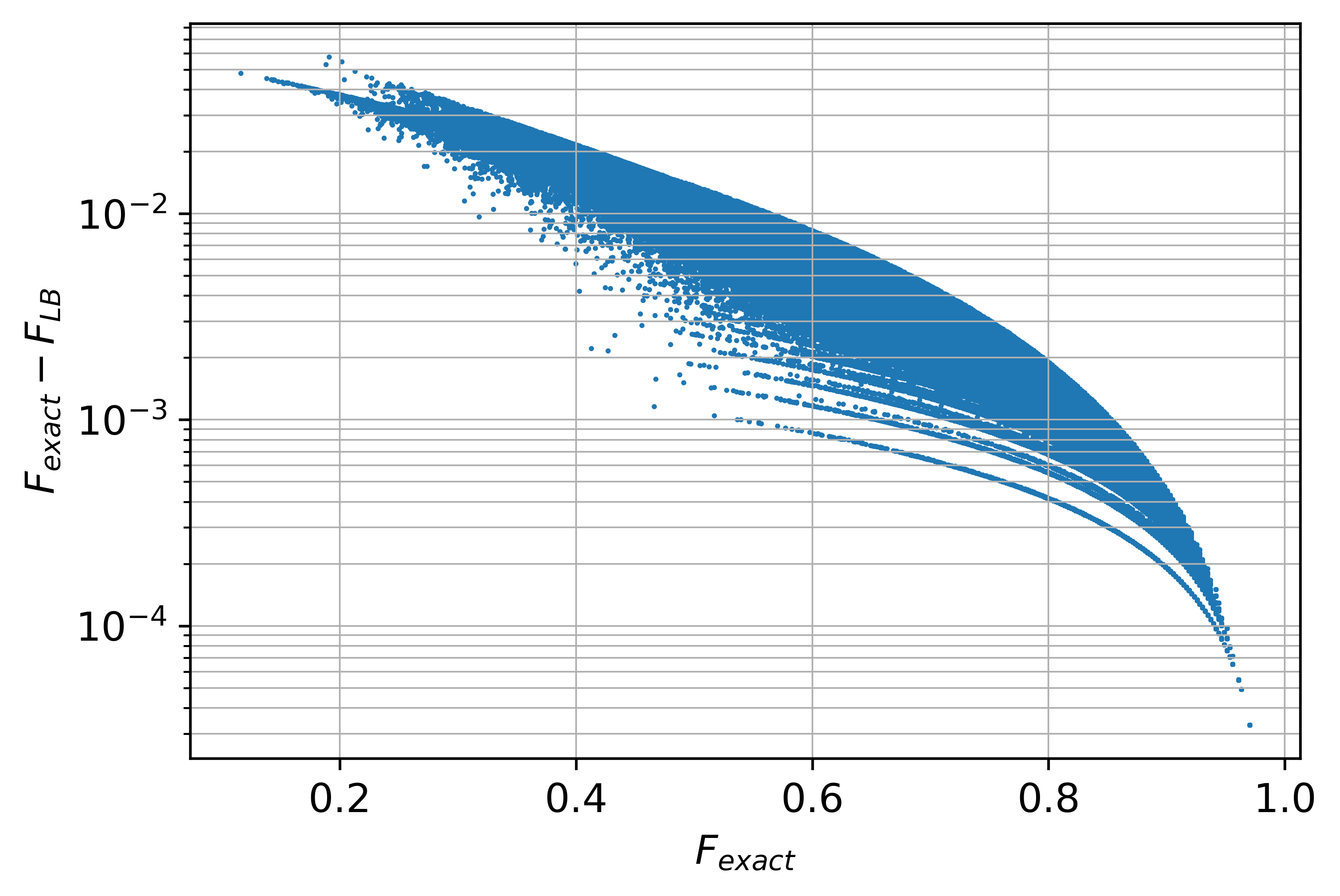}}\label{fig:f_errorapp}}
    \qquad
    \subfloat[\centering]
    {{\includegraphics[width =0.9\linewidth]{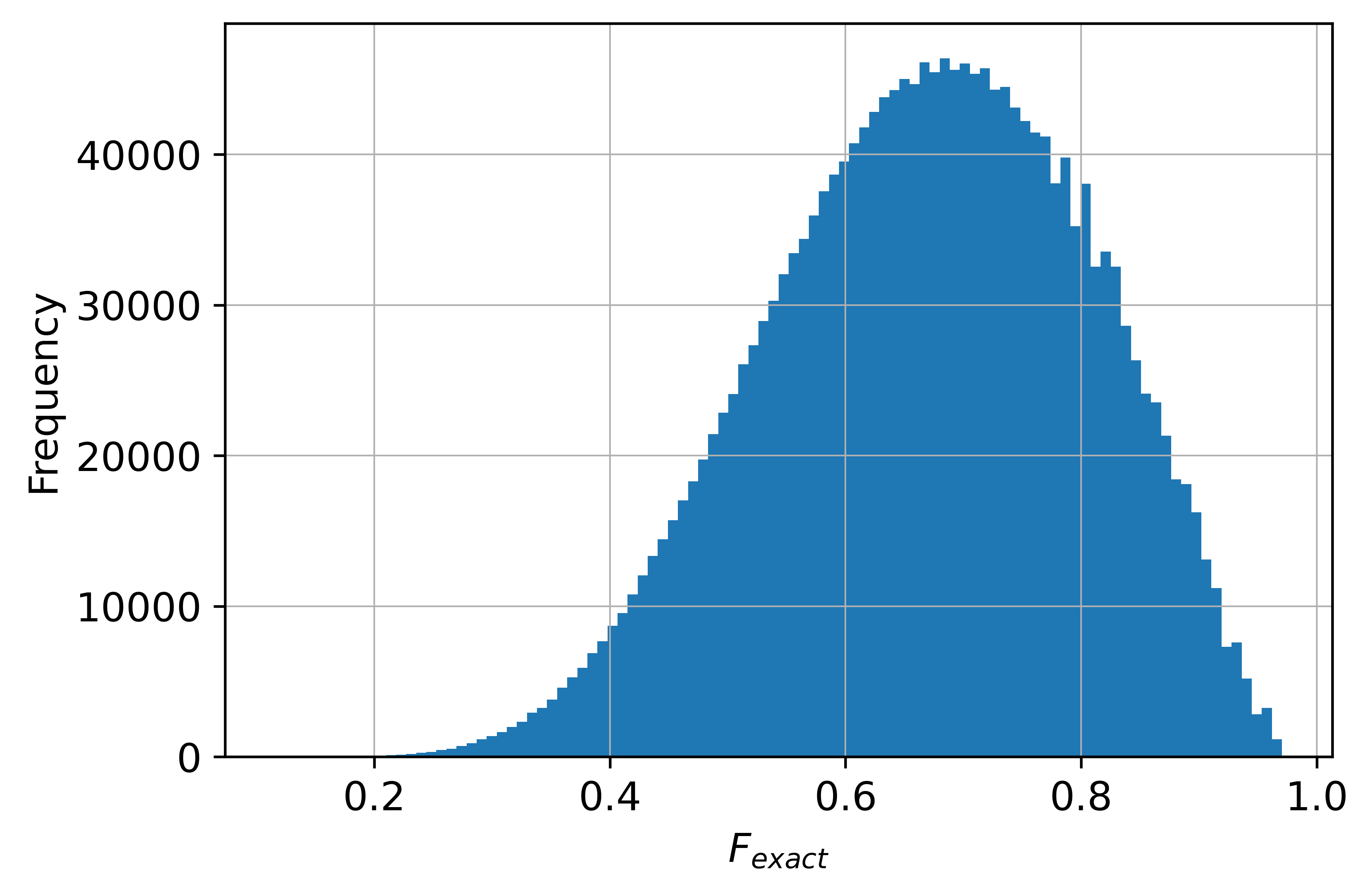}\label{fig:f_histapp}}}
    \qquad
    \caption{(a) Scatter plot of GHZ fidelity $\mathcal{F}_{exact}$ against $\mathcal{F}_{exact}- \mathcal{F}_{LB}$, which is the difference between exact fidelity and fidelity lower bound. (b) Histogram of $\mathcal{F}_{exact}$ for GHZ states generated in (a).}
    \label{fig:f_error_hist}
\end{figure}

Fig.~\mbox{\ref{fig:f_error_hist}} shows the fidelity of the $2.0 \times 10^6$ GHZ states simulated for Fig.~\mbox{\ref{fig:jointDR_F}}. Fig.~\mbox{\ref{fig:f_errorapp}} shows the difference between the lower bound $\mathcal{F}_{LB}$ and the exact fidelity $\mathcal{F}_{exact}$, across different $\mathcal{F}_{exact}$ values. Here $\mathcal{F}_{LB}$ was calculated as described in Eq. (\mbox{\ref{eq:fi_3}}), using the fidelity of Werner states generated after entanglement swapping. The value of $\mathcal{F}_{exact}$ was calculated using NetSquid \mbox{\cite{Coopmans2021netsquid}}. On the y-axis we see $\mathcal{F}_{exact}-\mathcal{F}_{LB}$ is strictly positive for all data points, further justifying the claim that $\mathcal{F}_{LB}$ is a lower bound. Note that $\mathcal{F}_{LB}$ is only used to calculate the routing solution, fidelity values shown are the exact fidelity $\mathcal{F}_{exact}$ unless otherwise specified. The average difference over the entire dataset was 0.0057 (1.07\%) and for the GHZ states of interest ($\mathcal{F}>1/2$), we see $\mathcal{F}_{exact}-\mathcal{F}_{LB}$ has values in the range $2 \times 10^{-2}-10^{-5}$. 
We find that $\mathcal{F}_{LB}$ is a good approximation when the contribution to fidelity of the higher order error terms is small. We found this to be true for the majority of GHZ states simulated and in particular for the GHZ states of interest (i.e. $\mathcal{F}>1/2$).

\section{Fusion based multi-path routing} \label{app:B}

Multi-path routing protocols have also been developed where each node performs fusion on all locally held Bell states via a GHZ-basis projective measurement. Such works can operate without requiring global link-state information and have been shown to achieve high distribution rates for distributing entangled states between two users \mbox{\cite{patil2020GHZFuse,clayton2024efficient,distillsinglet,QEP,acin2007entanglement}}. However, we found that such approaches reduced the achievable fidelity when the Bell states distributed in the network were noisy. In prior works on multi-path routing, Bell states are assumed to be either ideal \mbox{\cite{sutcliffetqe}}, error corrected 
\mbox{\cite{patil2020GHZFuse,clayton2024efficient}}, or distributed in specific states that enable singlet conversion to be used to produce ideal Bell states before fusion \mbox{\cite{distillsinglet,QEP}}. These approaches are not generally feasible with the network model considered due to the limited device capabilities (pre-error correction) and more general (depolarising) noise model, respectively.

We examined the performance of such fusion based approach by modelling two protocols, termed "mp-fuse-global" and "mp-fuse-local", that extend the fusion-based approach presented in \mbox{\cite{patil2020GHZFuse}} to consider the multi-user case. The protocols operate as follows:

\begin{enumerate}
    \item Attempt Bell state generation over all edges in $G$.
    \item Each \textit{non-user} node performs fusion (an n-qubit GHZ basis measurement \mbox{\cite{patil2020GHZFuse}}) between the qubits held of each successfully generated Bell state. This measures out all qubits at each of these nodes
    \item Each user node performs entanglement fusion (the 2-qubit operation as described in Section~\mbox{\ref{sec:gen}}) such that a single qubit remains unmeasured.
    \item Any non-user node with edge degree 1 in the link-state graph $G^\prime$ performs an $X$-basis measurement to remove the qubit from the shared GHZ state.
    \item Classical communication and local corrections (single-qubit Paulis) must be performed on the remaining qubits, with the exact operations depending on the measurement outcomes.
    \item After all such operations, a GHZ state has been successfully distributed between the user nodes. 
\end{enumerate}

The two fusion-based protocols differ in their requirements for global link-state information $G^\prime$. The protocol "mp-fuse-global" requires central knowledge of the global link-state and only executes the fusion operation once it has been confirmed centrally that all users are in the same connected component in $G^\prime$. This is a necessary and sufficient condition for a successful GHZ state distribution. The protocol \mbox{"mp-fuse-local"} performs the fusion operations after $Q_c$ rounds of entanglement link generation, without centralised knowledge of the global link-state information ($G^\prime$). As such, if any user is disconnected from the others in $G^\prime$ when the fusion is performed, this approach will fail to distribute the required GHZ state between all users and the whole operation must be repeated from step 1.

The distribution rate and fidelity of the fusion-based approaches is compared to the other protocols in Fig.~\mbox{\ref{fig:fusion}}, which extends the results shown in Fig.~\mbox{\ref{fig:jointDR_F}} Note that the fidelity value shown here is the fidelity lower bound $\mathcal{F}_{LB}$ described in Eq.~\mbox{(\ref{eq:fi})}. Fig.~\mbox{\ref{fig:fusion}} shows that mp-t and mp-fuse-global achieve the same distribution rate, but mp-fuse-global achieves worse fidelity. If the Bell states are noisy, the fidelity of a GHZ state generated from multiple Bell states (fusion or entanglement swapping) degrades with the number of Bell states required. For our protocols (mp-s or mp-t), the fidelity scales with the number of edges in the routing solution $O(\mathcal{F}_e^{|R|})$. For a fusion-based approach, the number of Bell states scales with the number of edges $|C|$ in the whole connected component $O(\mathcal{F}_e^{|C|})$ (connected component in $G^\prime$ that contains the users). For instances of $G^\prime$ for which routing is successful, the size of the connected component $C$ can be significantly larger than the size of the routing solution $R$, leading to significantly lower fidelities.

A key benefit of fusion-based approaches is that they can operate using only local link-state information. However, we see that the mp-fuse-local protocol achieves a significantly lower distribution rate compared to mp-fuse-global. Operating with only local link-state knowledge means that fusion operations will be performed regardless of the state of $G^\prime$. If fusion is performed while the users $S$ are disconnected in $G^\prime$, performing fusion will fail to distribute the required GHZ state. The GHZ state distributed will only be shared between users which are in the same connected component in $G^\prime$. If failure occurs, the protocol must reattempt GHZ state distribution, after generating a new set of Bell states. Such fusion-based approaches (using local link-state knowledge) perform well when Bell states are error corrected and can be generated with a probability above the percolation threshold.

\begin{figure}[h!tb]
    \centering
    \includegraphics[width=0.9\linewidth]{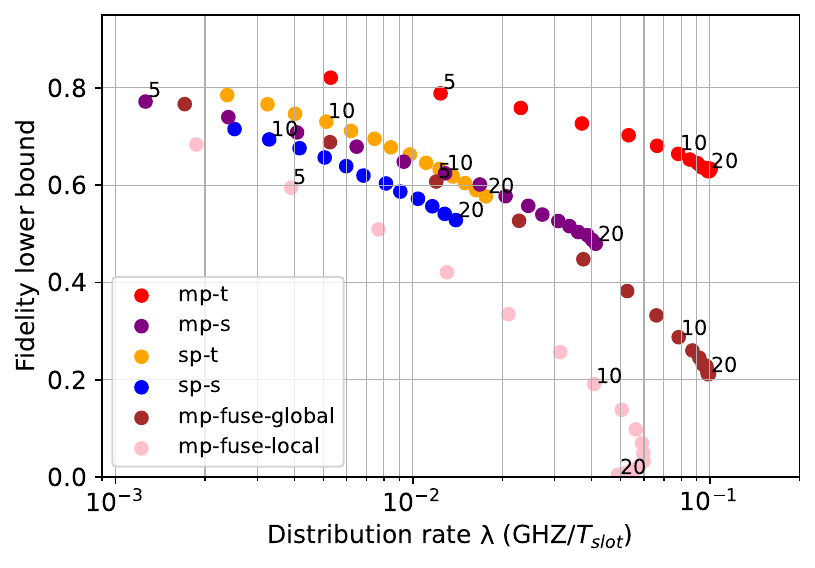}
    \caption{The protocols described in Section \ref{sec:protocols} are compared to two variants of a fusion based protocol mp-fuse-global and mp-fuse-local (that require either global or local link-state information).}
    \label{fig:fusion}
\end{figure}

\section{Sensitivity to entanglement link generation probability} 
\label{app:res_app}
The protocols were simulated for entanglement link generation probabilities $p = 0.2,0.3$ using otherwise identical assumptions as in Section \mbox{\ref{sec:res1}}. However, the results are averaged over 60 sets of users rather than 100. The entanglement link generation probability $p$ was adjusted to $p=0.2$ for Fig.~\ref{fig:app_1} and $p=0.3$ for Fig.~\ref{fig:app_2}. In Fig.~\ref{fig:app_1} the results show that when $p=0.2$, largely consistent with the results seen in Section~\mbox{\ref{sec:res1}}. The multi-path routing protocols provided up to a $9.5 \times$ higher distribution rate and $30\%$ higher GHZ state fidelity.
Fig.~\mbox{\ref{fig:app_2}} shows that for $p=0.3$ the maximum rate and fidelity improvements were $\times 7.6$ and $18\%$ respectively. The benefits of multi-path routing reduce as $p$ increases, as all protocols achieve higher distribution rates and fidelities when $p$ is higher.
Of the protocols, the highest absolute fidelity will be achieved by the sp-t protocol with $Q_c=1$. This uses the least cost routing solution with all entanglement links discarded at the end of each timeslot. Any multi-path routing solution will use at least as many entanglement links, hence reducing the average GHZ state fidelity. However, a cutoff of $Q_c=1$ may not be practical in near-term quantum networks. For $Q_c>1$, the multi-path protocols can achieve a significant improvement in distribution rate and fidelity.

\begin{figure}[h!tbp]
  \centering
  \begin{subfigure}[b]{0.9\linewidth}
    \includegraphics[width = 0.9\linewidth]{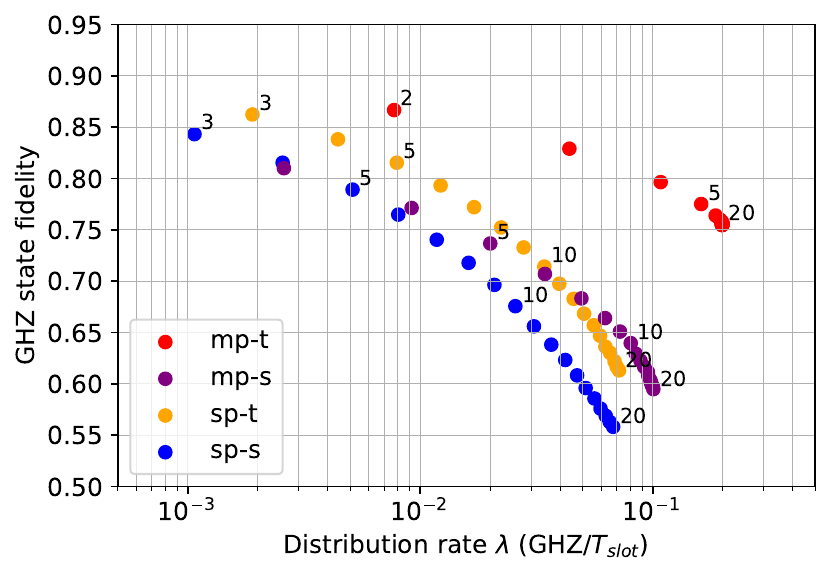}
    \caption{}
    \label{fig:app_1}
  \end{subfigure}
  \hfill
  \begin{subfigure}[b]{0.9\linewidth}
    \includegraphics[width = 0.9\linewidth]{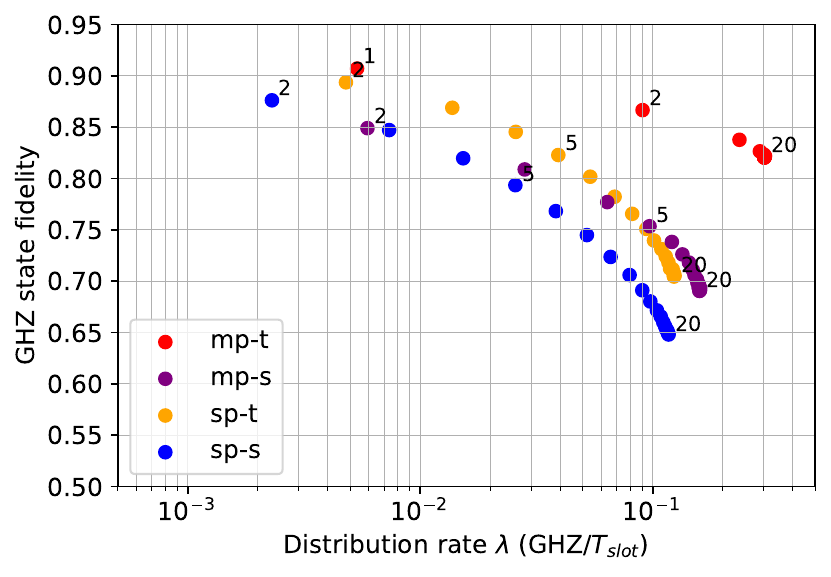}
    \caption{}
    \label{fig:app_2}
  \end{subfigure}
  \caption{Distribution rate against fidelity for the protocols simulated with $Q_c \in [1,20]$ in a $ 6\times 6$ grid graph. (a) shows the performance when $p=0.2$ and (b) with $p=0.3$.}
  \label{fig:app}
\end{figure}

\EOD
 
\end{document}